\documentclass[12pt,amssymb]{article} %Latex2e

\usepackage{amssymb}
\usepackage[dvips]{graphicx}
\newcommand{\newsection}[1]{
\addtocounter{section}{1} \setcounter{equation}{0}
\setcounter{subsection}{0} \addcontentsline{toc}{section}{\protect
\numberline{\arabic{section}}{{\rm #1}}} \vglue .6cm \pagebreak[3]
\noindent{\bf  \thesection. #1}\nopagebreak[4]\par\vskip .3cm}
\newcommand{\newsubsection}[1]{
\addtocounter{subsection}{1}
\addcontentsline{toc}{subsection}{\protect
\numberline{\arabic{section}.\arabic{subsection}}{#1}} \vglue .4cm
\pagebreak[3] \noindent{\it \thesubsection.
#1}\nopagebreak[4]\par\vskip .3cm}

\renewcommand{\theequation}{\thesection.\arabic{equation}}

\newlength{\extraspace}
\newlength{\extraspaces}
\newcounter{dummy}
\newcommand{\bc}{\begin{center}}
\newcommand{\ec}{\end{center}}
\newcommand{\be}{\begin{equation}
\addtolength{\abovedisplayskip}{\extraspaces}
\addtolength{\belowdisplayskip}{\extraspaces}
\addtolength{\abovedisplayshortskip}{\extraspace}
\addtolength{\belowdisplayshortskip}{\extraspace}}
\newcommand{\ee}{\end{equation}}

\newcommand{\ba}{\begin{eqnarray}
\addtolength{\abovedisplayskip}{\extraspaces}
\addtolength{\belowdisplayskip}{\extraspaces}
\addtolength{\abovedisplayshortskip}{\extraspace}
\addtolength{\belowdisplayshortskip}{\extraspace}}
\newcommand{\ea}{\end{eqnarray}}

\newcommand{\is}{& \!\! = \!\! &}
\newcommand{\ban}{\begin{eqnarray*}
\addtolength{\abovedisplayskip}{\extraspaces}
\addtolength{\belowdisplayskip}{\extraspaces}
\addtolength{\abovedisplayshortskip}{\extraspace}
\addtolength{\belowdisplayshortskip}{\extraspace}}
\newcommand{\ean}{\end{eqnarray*}}
\newcommand{\baa}{
\addtocounter{equation}{1} \setcounter{dummy}{\value{equation}}
\setcounter{equation}{0}
\renewcommand{\theequation}{\thesection.\arabic{dummy}\alph{equation}}
\begin{eqnarray}
\addtolength{\abovedisplayskip}{\extraspaces}
\addtolength{\belowdisplayskip}{\extraspaces}
\addtolength{\abovedisplayshortskip}{\extraspace}
\addtolength{\belowdisplayshortskip}{\extraspace}}
\newcommand{\eaa}{
\end{eqnarray}
\setcounter{equation}{\value{dummy}}
\renewcommand{\theequation}{\thesection.\arabic{equation}}}

\newcommand{\vev}[1]{\left\langle #1\right\rangle}

\newcommand{\half}{\frac{1}{2}}
\newcommand{\del}{\partial}
\newcommand{\eol}{\nonumber \\}

\newcommand{\cO}{{\cal O}}
\newcommand{\Ext}{{\rm Ext}}
\newcommand{\Hom}{{\rm Hom}}

\begin{document}

\begin{flushright}
December 2005\\
{\tt hep-th/yymmnnn}\\
PUPT-2186
\end{flushright}
\vspace{2cm}

\thispagestyle{empty}

\begin{center}
{\Large\bf  Parameter Space of Quiver Gauge Theories
 \\[13mm] }

{\sc  M.~Wijnholt}\\[2.5mm]
{\it Physics Department, Jadwin Hall, Princeton University\\
Princeton, NJ 08544}\\
[30mm]

{\sc Abstract}

\end{center}

\noindent Placing a set of branes at a Calabi-Yau singularity
 leads to an ${\cal{N}}=1$ quiver gauge theory. We analyze F-term deformations
 of such gauge theories. A generic deformation can be obtained by making the Calabi-Yau
 non-commutative. We discuss non-commutative generalisations of well-known singularities
 such as the Del Pezzo singularities and the conifold.

 We also introduce new techniques
 for deriving superpotentials, based on quivers with ghosts and a notion of generalised
 Seiberg duality. The curious gauge structure of quivers with ghosts is most naturally
 described using the BV formalism.
 Finally we suggest a new approach to Seiberg
 duality by adding fields and ghost-fields whose effects cancel
 each other.

  \vfill

\newpage

\renewcommand{\Large}{\normalsize}

\tableofcontents

\newsection{Parameter space of quiver gauge theories}

One of the most reliable ways to engineer a gauge theory from
string theory is by placing a set of D-branes in some background
geometry. If we require the gauge theory to be four-dimensional
with ${\cal {N}}=1$ supersymmetry, then up to dualities one
typically has to look at D-branes filling four flat dimensions and
wrapped on (possibly collapsed) cycles in a Calabi-Yau three-fold.
Embedding a gauge theory into string theory is relevant for at
least two of the main threads of research: it generates examples
of the ADS/CFT correspondence, and it is a first step towards
bottom-up string phenomenology. Apart from this, the gauge theory
is closely tied to the Calabi-Yau geometry, and there are amusing
relations with modern areas of mathematics.

In this article we will focus on the gauge theory one obtains from
a set of $N$ D3-branes located at a Calabi-Yau three-fold
singularity in type IIb string theory. The theories one obtains
this way are of quiver type, and for $N>1$ are believed to flow to
interesting interacting conformal field theories. For applications
to either ADS/CFT or phenomenology, one would like to understand
the possible deformations of the gauge theory.

By ADS/CFT intuition it is tempting to believe that small
deformations of the gauge theory can still be realised after
embedding in string theory. This is particularly clear when the
theory is conformal and the deformations are marginal.
Nevertheless, if one examines the quiver gauge theory, the number
of deformations is larger than the number of conventional
geometric deformations of the local Calabi-Yau geometry. So the
puzzle is how to identify the full parameter space of the quiver
in string theory.

We will examine a number of well-known Calabi-Yau singularities
and account for all the marginal deformations that can be
understood as F-term data (i.e. superpotential deformations). Some
of these deformations can be understood as conventional complex
structure deformations of the Calabi-Yau, and were previously
investigated by the author in \cite{Wijnholt:2002qz}. Here we find
that all the remaining deformations of these quiver theories can
be understood as non-commutative deformations of the
Calabi-Yau.\footnote{In particular we find the missing
deformations of \cite{Verlinde:2005jr}.} We emphasize that the
four-dimensional gauge theory living on the branes is a
conventional commutative gauge theory.

As in our previous work, in order to uncover the map between the
gauge theory parameter space and the Calabi-Yau parameter space,
we will need to make use of the general technique of exceptional
collections. Other approaches that the author is aware of are not
flexible enough to deal with deformations. Another complication is
that in the presence of non-commutative deformations, the moduli
space of the quiver theory for a single D3-brane is not the
Calabi-Yau itself.

Another topic we address here is the effect of certain braiding
operations on the quiver diagram. It is known
\cite{Cachazo:2001sg} that a subset of such operations can be
understood as Seiberg duality on the gauge theory. However there
are more general operations (`generalised Seiberg dualities')
which do not have an immediate gauge theory interpretation.
Building on some unpublished work with Cachazo, Katz and Vafa
\cite{oldquiver} we discuss how to deal with the resulting
quivers. The more general quivers one obtains this way can be
thought of as quivers with ghosts, and this leads to a consistent
way of manipulating them. Our point of view here is not that these
manipulations can be carried out in field theory -- indeed we do
not know how to associate a sensible gauge theory to a quiver with
ghosts -- however, it is that these manipulations make sense at
the level of F-terms and can be used as a technique for computing
topological data such as superpotentials in physical quiver gauge
theories.

Relations between quivers and non-commutative Calabi-Yau spaces have
previously been pursued in the series of papers
\cite{Berenstein:2000ux,Berenstein:2000te,Berenstein:2001jr,She:2005qq}.
Other aspects of exceptional collections have recently been explored
in \cite{Herzog:2005sy,Bergman:2005kv,2005math......6166A}. A word
on notation: when we write superpotentials, the overall trace will
be implicit.

\newsection{Large volume construction of quiver theories}

\newsubsection{Topological amplitudes}

We are interested in the low energy gauge theory for a set of
branes placed at a Calabi-Yau singularity in type IIb string
theory. It is generally believed that this gauge theory can be
described in terms of a basis of `fractional branes' which depend
on the singularity. There is no general proof of this statement
because the conformal field theory is typically not under complete
control, but many cross-checks have been made and the fractional
brane picture holds up rather well.

So we assume that there is a set of boundary states $\{ F_1
,\ldots, F_n\}$ `localised' at the singularity, with the following
properties:
\begin{itemize}
\item the RR charge vectors (which describe the coupling to RR fields) form a basis for the homology lattice
of vanishing cycles;
\item they all break the same half of the 8 supercharges, i.e. they are mutually BPS;
\item they are `irreducible' and other possible branes can be expressed as bound states of the fractional branes.
\end{itemize}
For a discussion of the last item, see \cite{Douglas:2000gi}.

Suppose then we want to describe the worldvolume theory of some
set of branes. For convenience we will take the case of a
D3-brane, which corresponds to some boundary state $F_p$ on the
Calabi-Yau, and let us denote the RR charge vector by `ch.' Then
we can first decompose $F_p$ into the $F_i$ at the level of
homology:
\be
{\rm ch}(F_p) = \sum_{i=1}^n n_i \,{\rm ch}(F_i) .
\ee
The massless fields arising from open string modes are easy to
recognise. From open strings stretching between the $|n_i|$
fractional branes of type $i$ one expects a vector multiplet in
the adjoint of $U(|n_i|)$. Also, for each `intersection' of two
fractional branes one expects a chiral multiplet. By
`intersection' we mean the intersection of the vanishing cycles
associated with a fractional brane according to its charge vector.
Even in our case where we only have even cycles, such
intersections should be counted with a sign (as is also required
in order to be consistent with mirror symmetry). Depending on this
sign, one gets a chiral multiplet in the anti-fundamental of $F_i$
and the fundamental of $F_j$ or reversely.

This minimal amount of data determines the massless field content and therefore a large part of the low energy theory.
This data, as is well known, can be summarised in a quiver diagram. To fix the parameters in this low energy gauge theory
requires one to compute a finite set of string amplitudes and compare with the corresponding amplitudes of the effective gauge theory.
Thus our main concern is to find a good basis of fractional branes. Unfortunately except for the case of orbifolds of flat space
(i.e. free field theory) one is unable to do that.

There is a trick however if we restrict ourself to a topological subsector of the full open string theory. In our setting
this is  the (open
string) B-model. The matter part of a string vertex operator is composed of a four-dimensional part and an internal
six dimensional part that lives on the Calabi-Yau. A certain class of string amplitudes can be computed in the topologically
twisted theory. From the gauge theory point of view, this is the set of amplitudes that can be calculated just from the F-terms
without using any information from the D-terms.

The beauty of this class of amplitudes is that they do not depend
on the (complexified) K\"ahler parameters of the Calabi-Yau, since
those would only affect the D-terms. From the point of view of the
topological BRST operator, K\"ahler deformations are exact.
Therefore we can change the K\"ahler parameters and go to a point
in moduli space where we do know the conformal field theory. Such
a point is given by the large volume limit, where we can use the
non-linear sigma model. In this limit, one can describe the
fractional branes as certain exceptional collections of sheaves
localised on the collapsing cycles.\footnote{Actually it is clear
that this needs to be slightly generalized, for instance
condensing some fields in a quiver obtained from a three-block
exceptional collection does not give rise to another exceptional
collection. We will omit such subtleties from the discussion.}

In the nl$\sigma$m description, the massless open string modes are
counted by certain cohomology groups, the global Ext groups. Thus,
given two sheaves  $F_i,F_j$ localised on collapsing cycles, we
should first extend the sheaves to $i_*F_i, i_* F_j$ on the
Calabi-Yau three-fold (where $i$ is the embedding of the collapsing
cycles into the Calabi-Yau three-fold), and then compute\footnote{We
will often drop the push-forward symbol ``$i_*$'' in the remainder
of this section, to simplify notation.}
\be\label{openstring}
\Ext^p(i_*F_i,i_* F_j)
\ee
The grade $p$ is called the ghost number of a topological vertex
operator (it is however in the matter sector in the full ten
dimensional string theory, which uses a different ghost number
symmetry).

As argued in \cite{Douglas:2000gi}, the correct large volume
description of fractional branes is typically not just a set of
sheaves. We should also do some spectral flow on the boundary
conditions, using the $U(1)$ generator of the worldsheet $N=2$
algebra. Unless one takes this into account, one finds that the
ghost number of a vertex operator (which is just the charge under
this $U(1)$) may not have the same value in the large volume
limit. In order to account for this, one embeds the sheaves in the
derived category, where the spectral flow we need to repair the
ghost number is interpreted as a shift in the position in the
complex. Changing the vertex operators by spectral flow is
strictly not needed in that the correlation functions in our
context are only changed by a trivial factor, but it is
nevertheless useful to keep track of it. Spectral flow will be
indicated with the conventional derived category notation, eg.
$F[k]$ denotes $F$ with $k$ units of spectral flow applied.
The ghost number of
\be
V \, \in \, \Ext^p(E[q],F[r])
\ee
is $N_{\rm gh} = p - q + r$. In the
following we will assume that the appropriate shifts have been
made in (\ref{openstring}).

It is a well known fact that the usual physical vertex operators sit
at ghost number one, but in principle one can have `ghosts', i.e.
BRST cohomology classes at ghost numbers different from one. The
open string field theory for the B-model is of Chern-Simons type
\cite{Witten:1992fb}, and the appearance of many ghosts is quite
typical if one quantizes such a theory. As we review momentarily,
operators at ghost number zero are associated with symmetries
(`boundary ground ring'), operators at ghost number minus one with
`ghosts for ghosts' (symmetries among symmetries), etc.

We could also get vertex operators with ghost number $p>1$. These
are called the anti-fields. They have the interpretation of
obstructions to the deformations, obstructions for the obstructions,
and so on. There is a pairing between vertex operators of ghost
number $p$ and ghost number $3-p$ given by the disk two-point
function (which evaluates to the Serre duality pairing). Given a
vertex operator of ghost number $p>1$, it is more natural for us to
consider its dual under this duality pairing. For instance the dual
of a ghost number $p=2$ operator has ghost number $p=1$, and thus it
can be interpreted as a deformation. In our context this corresponds
to moving the brane away from the collapsing cycle in the
non-compact direction. After GSO projection, an operator of ghost
number $p=1$ gives rise to a chiral field in four dimensions, and
its dual operator of ghost number $p=2$ gives rise to the conjugate
anti-chiral field. This is also familiar from heterotic model
building on a Calabi-Yau \cite{GSW}.

Given a set of $V_i$ vertex operators of ghost number one, with
$V_i \in \Ext^1( F_i, F_{i+1})$ (and assuming $F_{n+1} = F_1$ then
we can define a disk amplitude as
\be\label{diskamp}
\vev{ V_1(\infty) V_2(0) V_3(1) \int_1^{y_4} V_4 \int_{y_4}^{y_5} V_5 \ldots \int_{y_{n-1}}^\infty V_n }
\ee
In the low energy gauge theory we get  the analogous tree level
amplitude to be proportional to a certain coefficient in the
superpotential, namely the coefficient of
\be
{\rm Tr}\  \int d^2 \theta\,  \Phi_1 \Phi_2 \Phi_3 \ldots \Phi_n .
\ee
(since we cannot use the K\"ahler terms and since there is no mass
term in the superpotential, it is impossible to build an $n$-point
Feynman diagram by contracting lower-point vertices). Thus the
amplitudes (\ref{diskamp}) compute coefficients in the
superpotential.

\newsubsection{Ghost number zero operators}

%%%%
 \begin{figure}[th]
\begin{center}
        %\resizebox{\textwidth}{!}{
            \scalebox{0.7}{
               \includegraphics[width=\textwidth]{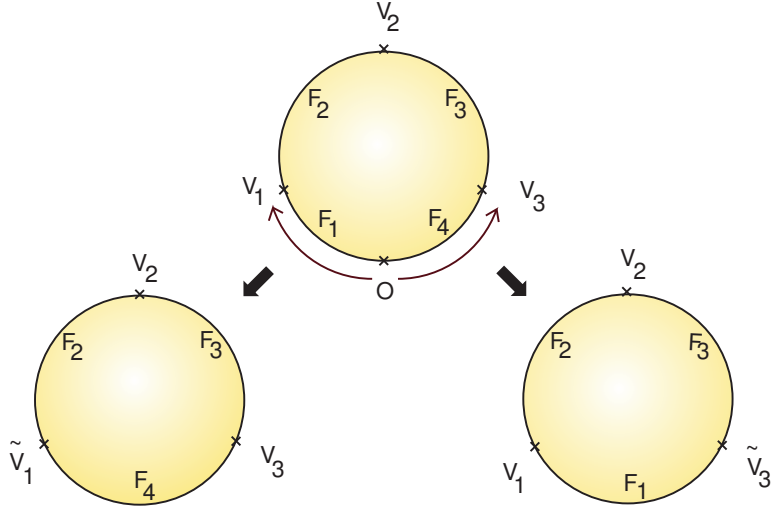}
               }
\end{center}
\vspace{-.5cm}
\caption{ \it Ghost number zero operators can be used to relate disk amplitudes.}\label{disksymm}
 \end{figure}

Now let us try to understand the role of vertex operators of ghost
number different from one. Suppose again that $V_i \in \Ext^1( F_i,
F_{i+1})$ and consider $O \in \Ext^0(F_4,F_1)$. Consider the
amplitude
\be
\vev{V_1(\infty) V_2(0) V_3(1) O(y) }
\ee
Note that since $O(y)$ has ghost number zero, it should not be
integrated over the boundary. Now the amplitude is independent of
the position of the ghost number zero operator (since $\del O = \{
Q, b_{-1} O \}$). So we can take the limit $y\to 1$, in which case
we get
\be
\lim_{y \to 1} V_3(1) O(y) = - V'_3(1) \in \Ext^1(F_3,F_1)
\ee
For chiral primaries there are no poles in the OPEs. Alternatively
we can take the limit $y \to \infty$, in which case we have
\be
\lim_{y\to \infty} O(y) V_1(\infty) = V'_1(\infty) \in \Ext^1(F_4,F_1).
\ee
Therefore we find
\be
\vev{V_1(\infty) V_2(0) V'_3(1)} + \vev{V'_1(\infty) V_2(0) V_3(1) }=0.
\ee
In other words, the ghost number zero operators generate relations
among the superpotential couplings. Such symmetries in turn
guarantee the existence of flat directions. Namely the
superpotential terms ${\rm Tr}(\Phi_1 \Phi_2 \Phi'_3 + \Phi'_1
\Phi_2 \Phi_3)$ are invariant under
\be\label{gauge}
 \delta \Phi'_1 = \Lambda \Phi_1, \qquad \delta \Phi'_3 = -\Phi_3 \Lambda.
 \ee
An expectation value for $\Lambda$ (which we may think of as the
four-dimensional partner of $O$) has no interpretation in the
D-brane system, it is purely a redundancy of the description.
Therefore we should mod out by such symmetries. If $F_1$ and $F_4$
correspond to identical boundary conditions, this is easy to
understand; in this case the transformations (\ref{gauge}) just
correspond to the non-abelian gauge transformations that arise when
you have a stack of identical branes on top of each other. However
we will see examples in which $F_1 \not = F_4$, and there is a
generator in $\Ext^0(F_4,F_1)$ but not in $\Ext^0(F_1,F_4)$. In that
case the ghost number zero operators do not generate a reductive Lie
algebra, i.e. a sum of simple and semi-simple algebras, but a
parabolic algebra, and it seems impossible to gauge it and preserve
CPT.

Even though we seem to be unable to associate a physical quiver
when we have parabolic symmetries, it will be convenient to
associate quiver diagrams to such exceptional collections and
manipulate them. Any such collection should contain all the
information about F-terms. For each parabolic generator we can
introduce a ghost field $\Lambda$ which is a chiral field except
with the opposite statistics.
 Because of the unusual statistics the corresponding arrow in the quiver diagram should
be reversed. Similar remarks apply to operators of ghost number
$p<0$. These correspond to ghosts-for ghosts, etc. Of course
cohomology classes of topological ghost number $p$ do not
necessarily correspond to cohomology classes of physical ghost
number $p$, since the ghost number grading in the 10D string
theory is different. For instance cohomology classes of ghost
number zero that live in an adjoint representation give rise to
physical vector multiplets. Our proposal here is to treat cohomology classes
that live in a bifundamental representation as having the same physical and
topological ghost number. We will see this is a
useful perspective, at least at the level of F-terms.

It has been suggested in the literature that bifundamentals
obtained from $\Ext^0$ cohomology classes should correspond to
tachyons. This is incompatible with the point of view taken here,
since only fields of the right ghost number can get expectation
values. In particular we wish to avoid giving expectation values
to gauge redundancies.

\newsubsection{Review of Seiberg duality and mutations}

%%%%
 \begin{figure}[th]
\begin{center}
        %\resizebox{\textwidth}{!}{
            \scalebox{0.4}{
               \includegraphics[width=\textwidth]{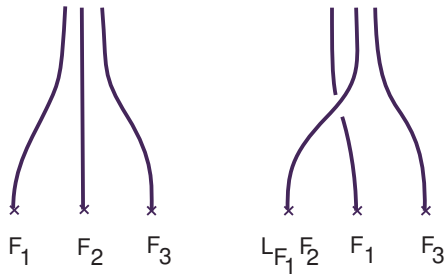}
               }
\end{center}
\vspace{-.5cm} \caption{ \it A braiding operation on the
collection of fractional branes. These pictures can be interpreted
in terms of D6 branes wrapped on Lagrangian cycles in the mirror
\cite{Hori:2000ck}.}\label{braiding}
 \end{figure}

We have assumed the existence of a set of boundary states $\{ F_1,
\ldots, F_n \}$ which gets mapped to an exceptional collection in
the large volume limit. However for any given singularity there are
infinitely many such collections. This is actually not completely
surprising, because so far we have only really defined the complex
structure of the local singularity, and all collections contain the
same holomorphic information. The existence of many collections for
a given singularity reflects the fact that there are many points in
the K\"ahler moduli space where the cycles are collapsed to zero
size. If we interpolate between such points, the basis of vanishing
cycles may undergo a Picard-Lefschetz monodromy. The collection $\{
F_1, \ldots, F_n\}$ comes with an ordering, and the effect of the
monodromy is that a sheaf may be moved to the left or to the right
in the collection. When a sheaf $F_i$ is moved to the left or to the
right, we end up with a new exceptional collection $\{ F_1, \ldots,
L_{F_{i-1}}F_i, F_{i-1}, F_{i+1}, \ldots, F_n \}$ or $\{ F_1,
\ldots, F_{i-1}, F_{i+1}, R_{F_{i+1}}F_i, \ldots, F_n\}$, as
indicated in figure \ref{braiding}.

The charge vector of the new sheaf is given by the characteristic
Picard-Lefschetz formula:
\be\label{PLmon} {\rm ch} (F_i) \to {\rm ch}(L_{F_{i-1}} F_i)
=\pm[{\rm ch} (F_i)-\chi(F_{i},F_{i-1}) {\rm ch}(F_{i-1})]. \ee
Such a monodromy arises around a locus in the moduli space where
the central charge of $F_{i-1}$ (i.e. its period) goes to zero.
 An action on sheaves which has the effect of
(\ref{PLmon}) on the charge vectors is called a mutation or a
braiding operation. A mutation turns one exceptional collection
into another, and (up to some `trivial' operations like tensoring
the whole collection with a line bundle) for the cases we consider
all exceptional collections may be related through a sequence of
mutations. However a Picard-Lefschetz monodromy is typically a
composition of a few mutations; not every individual mutation may
be realized as a monodromy in the K\"ahler moduli space.

Once we specify both the complex and the K\"ahler structure of the
local geometry the collection should be uniquely specified. The idea
is that an exceptional collection becomes valid if the corresponding
fractional branes become mutually supersymmetric, that is if the
periods of the fractional branes (which depend on the K\"ahler
moduli) line up in the complex plane and have the same phase.
Evidence for this picture has been given in
\cite{Douglas:2000qw,Aspinwall:2004mb}. Now suppose further that we
take a path in moduli space so that the absolute value of a period
of one of the fractional branes goes to zero. Then we expect to get
a new collection related by a Picard-Lefschetz monodromy and hence a
different quiver gauge theory. The gauge theory interpretation of
this is that the gauge coupling associated with the corresponding
node blows up, and we get a new quiver related by Seiberg duality to
the old one.

%%%%
 \begin{figure}[th]
\begin{center}
        %\resizebox{\textwidth}{!}{
            \scalebox{0.9}{
               \includegraphics[width=\textwidth]{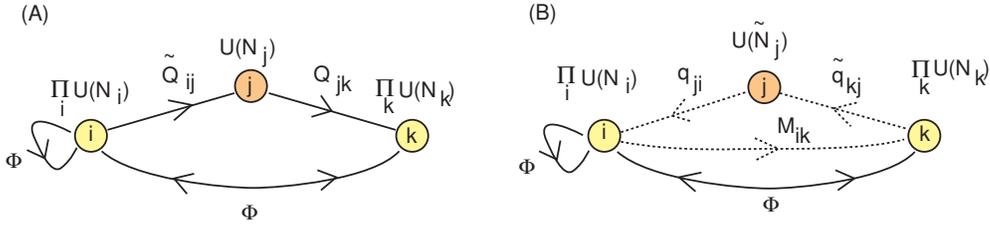}
               }
\end{center}
\vspace{-.5cm} \caption{ \it (A): Organising the nodes before
appkying a Seiberg duality. (B): The quarks $Q$ are replaced by
the dual quarks $q$ and the mesons $M = Q \tilde{Q}$, and the
gauge group is changed from $SU(N_c)$ to $SU(N_f -
N_c)$.}\label{SDual}
 \end{figure}

Now we can see why only a subset of mutations appears to be
realised through monodromies in the K\"ahler moduli space. Suppose
we want to do a Seiberg duality on node $j$ (see figure
\ref{SDual}). Let us organise the quiver so that all the incoming
arrows are all to the left of $j$, and all the outgoing arrows are
to the right of $j$. Then one can show \cite{Cachazo:2001sg} that
a Seiberg duality corresponds to a mutation by $j$ of either (1)
all the nodes to the left, or (2) all the nodes on the
right.\footnote{There may also be nodes which are not connected to
node $j$ by an arrow; such nodes are not changed under a Seiberg
duality.} If we decide to perform a mutation on only a few of the
nodes on the left or on the right, then we end up with a quiver
with ghosts, for which there does not seem to be a physical
interpretation. Nevertheless as we have explained it is possible
at the level of F-terms to make sense out of quivers with ghosts,
and all such quivers are related through mutations which are not
Seiberg dualities. We can therefore view mutations as a
`generalised Seiberg duality.' Since there typically are quivers
with ghosts that are very easy to calculate, then we can use
generalised Seiberg dualities as a technique for deriving ordinary
physical quiver gauge theories without ghosts. This will be
explained  in section 4.

\newsubsection{Holomorphic deformations of quiver theories}

A quiver gauge theory admits a large number of deformations. Here we
are interested in deformations of the F-terms, i.e. ratios of
superpotential couplings that are invariant under field
redefinitions. Such deformations should be given by perturbing the
closed string B-model by vertex operators of ghost number $2$. The
BRST cohomology at ghost number $2$ lives in the following
cohomology groups \cite{Witten:1991zz}:
\be
\sum_{i+j=2} H^i(X,\Lambda^j T_X) = H^0(X,\Lambda^2 T_X) \oplus H^1(X,T_X) \oplus H^2(X,\cO_X)
\ee
that is, tensors of type $\mu^{ij}, \mu^i_{\bar{j}}$ or $\mu_{\bar{i}\bar{j}}$. The interpretation of
these deformations is as follows:
\begin{itemize}
    \item $H^1(X,T_X)$ counts classical complex structure deformations of $X$;
    \item $H^0(X,\Lambda^2 T_X)$ counts global holomorphic Poisson structures, in other words,
    non-commutative deformations (inverse $B$-fields);
    \item $H^2(X,\cO_X)$ counts `gerbe' deformations obtained by turning on a $B$-field
    with two anti-holomorphic indices.
\end{itemize}
We will further restrict ourselves to exactly marginal deformations
of the conformal theory living on $N$ D3-branes placed at the
singularity. Since the radial direction away from the singularity
has the interpretation of an energy scale, the scale invariance of
the SCFT means that the local geometry is that of a complex cone
over a compact complex surface. Marginal deformations will preserve
this structure, and so such deformations correspond to deformations
of the complex surface.\footnote{For ${\cal B}_1$ and ${\cal B}_2$
it appears that some of the operators we use to deform the
superpotential do not have $R$-charge exactly equal to two. This is
related to the fact that the parameter that simply rescales the
complex variables is not exactly the radial direction in the
Calabi-Yau metric in these examples. We thank Sergio Benvenuti for
pointing this out.}

It must be emphasized that there are plenty of other F-term
deformations that are not of this type, and that can become large
either close to the tip of the cone or very far away. For instance,
we could be interested in adding fractional branes which lead to
non-perturbative behaviour in the IR triggering an extremal
transition. Or we could be interested in adding relevant or
irrelevant terms to the superpotential, such as mass terms for the
adjoints in $N=4$ YM theory (i.e. the $N=1^*$ and $N=2^*$
deformations). All these cases are captured by the B-model,
generically on a generalized geometry. But here we will restrict
ourselves to scale invariant deformations.

The main class of examples that we consider in detail are the Del
Pezzo singularities. Recall that a Del Pezzo surface is either a
${\bf P}^2$ blown up at $k$ points (often denoted ${\cal B}_k$) or
${\bf P}^1 \times {\bf P}^1$ (often denoted as ${\bf F}_0$). On such
a surface $h^2(X,\cO_X)=h^{(0,2)}=0$ so we don't get any gerbe
deformations. On ${\cal B}_k$ we naively expect $2k$ complex
structure parameters (describing the position of $k$ points on ${\bf
P}^2$) and $10-k$ NC deformations (since $\Lambda^2 T_X$ is
isomorphic to the line bundle of cubic homogeneous polynomials that
vanish at the points that get blown up). Finally the group
$PGl(3,{\bf C})$ of holomorphic coordinate redefinitions kills 8 of
these parameters, so in total we expect $k+2$ deformations for
${\cal B}_k$. Similarly for ${\bf F}_0$ there are 3 deformations.
This agrees with the allowed number of deformations that one can
read off from the quiver diagram, as one can check easily.

\newpage

\newsection{Non-commutative singularities and deformations of superpotentials}

\newsubsection{Non-commutative deformations}

First we need to discuss some basic properties of non-commutative
algebraic geometry. See \cite{1999math.....10082S} for a more
rigorous review. The first case we will consider is ${\bf
C}^3/Z_3$ which has a collapsed ${\bf P}^2$. Suppose we have a
brane that wraps ${\bf P}^2$ and suppose we turn on a $B$-field
with purely holomorphic indices. We will extend the $B$-field to
the Calabi-Yau three-fold by making it independent of the radial
coordinate of the Calabi-Yau and its complex partner. Then $H =
dB$ is identically zero, and its inverse $\theta^{ij} \sim
(B^{-1})^{ij}$ is a section of $\Lambda^2 T_{{\bf P}^2}$. This
bundle has many holomorphic sections which we can use to deform
the sigma model. The general effect of turning on $\theta^{ij}$ is
to deform the left- and rightmoving $N=2$ algebra on the
worldsheet so that the left- and rightmoving complex structures
are no longer equal \cite{Kapustin:2003sg,Gates:1984nk}.
Geometrically this situation can be described using generalized
complex geometry \cite{Hitchin,Gualtieri}. For the purposes of
this paper we are interested in the effect of turning on
$\theta^{ij}$ on open strings.  Then we expect the coordinates on
${\bf P}^2$ to become non-commutative according to
\be [ x^i, x^j ] = \theta^{ij}(x). \ee
Here $x^i,x^j$ are local coordinates; eg. in the patch $z^3 \not =
0$ they are of the form $(z^3)^{-1} z^1, (z^3)^{-1}z^2$. If
$\theta^{ij}$ is holomorphic than this is a type of complex
structure deformation and should make an appearance in the
superpotential.

It will be convenient to express the commutation relations in
projective coordinates rather than local coordinates. It is known
that a generic NC structure on ${\bf P}^2$ can be put in the form
\cite{ATV}
\ba\label{sklyanin}
 \alpha z_1 z_2 + \beta z_2z_1 + \gamma z_3^2 = 0 \eol
\alpha z_2 z_3 + \beta z_3z_2 + \gamma z_1^2 = 0 \eol
\alpha z_3 z_1 + \beta z_1z_3 + \gamma z_2^2 = 0
\ea
which is known as an `elliptic algebra' or a `Sklyanin algebra.'
These equations are familiar from the F-term equations for the
Leigh-Strassler deformations of $N=4$ Yang-Mills theory
\cite{Leigh:1995ep}, which is indeed known to be related to
non-commutative deformations \cite{Berenstein:2000ux}. In fact the
Leigh-Strassler deformations are invariant under the trihedral
group $\Delta_{27}$,\footnote{We would like to thank Sergio
Benvenuti for pointing this out to us.} and when we orbifold by a
$Z_3$ subgroup to get ${\bf C}^3/Z_3$ the Leigh-Strassler
deformations descend to the NC deformations of the quotient.
Nevertheless it will be useful to proceed with our point of view
because it can easily be extended to non-orbifold singularities.

When writing homogeneous equations in non-commutative coordinates,
one can assign an integer grade to a coordinate depending on which
position in a monomial it appears. In this subsection, we will
denote a coordinate in the first position as $A^i$ and a
coordinate in the second position as $B^i$. It will be evident
shortly why this is a useful thing to do.

Then we may rewrite the above equations (\ref{sklyanin})
as
\be
\left(%
\begin{array}{ccc}
  \beta A^2 & \alpha A^1 & \gamma A^3 \\
  \gamma A^1 & \beta A^3 & \alpha A^2 \\
  \alpha A^3 & \gamma A^2 & \beta A^1 \\
\end{array}%
\right)
\cdot
\left(%
\begin{array}{c}
  B^1 \\
  B^2 \\
  B^3 \\
\end{array}%
\right)
\equiv f_{ijk}A^i B^j = 0
\ee
These equations determine a variety in ${\bf P}_A^2 \times {\bf
P}_B^2$ which one can think of as the graph of a linear
isomorphism of a certain elliptic curve. The elliptic curve is
given by $A^i \in {\bf P}_A^2$ det$(f_{ijk} A^i)=0$, which gives
\be\label{ellcurve}
\alpha\beta\gamma((A^1)^3 +(A^2)^3 + (A^3)^3) - (\alpha^3 + \beta^3 + \gamma^3)A^1 A^2 A^3=0
\ee
If $A^i$ lies on this elliptic curve, the matrix $f_{ijk} A^i$ has
rank 2, so it has a one-dimensonal kernel spanned by some vector
$B_A^i \in {\bf P}_B^2$.  Note that $B_A^i$ must lie on the
elliptic curve det$(f_{ijk} B^j) = 0$, i.e.
\be
\alpha\beta\gamma((B^1)^3 +(B^2)^3 + (B^3)^3) - (\alpha^3 + \beta^3 + \gamma^3)B^1 B^2 B^3=0
\ee
Therefore, $f_{ijk}$ determines an automorphism of the elliptic
curve, given by
\be
\sigma(A^i) = B_A^i .
\ee
To abbreviate the notation, we will often write $\sigma(A^i) =
(A^\sigma)^i$.

The elliptic curve and the automorphism (which can be thought of
as translating by some point $\eta$ on the curve) completely
characterise the NC structure on ${\bf P}^2$. Clearly we have for
any point $p$ on the elliptic curve (\ref{ellcurve})
\be
f_{ijk} p^i (p^\sigma)^j = 0
\ee
Thus intuitively the NC structure degenerates along the elliptic
curve we have discussed, and we can think of this curve as an
embedded commutative curve. The more precise statement is that the
twisted homogeneous coordinate ring of the curve is equivalent to
a commutative ring, in that it has the same modules
\cite{1999math.....10082S}.

It is possible to give a more explicit parametrisation of
$p^\sigma$ for general $p$ by uniformizing the elliptic curve
using $\theta$-functions. See \cite{Odesskii:2002aa} for details.

\newsubsection{The projective plane}

%%%%
 \begin{figure}[th]
\begin{center}
        %\resizebox{\textwidth}{!}{
            \scalebox{0.8}{
               \includegraphics[width=\textwidth]{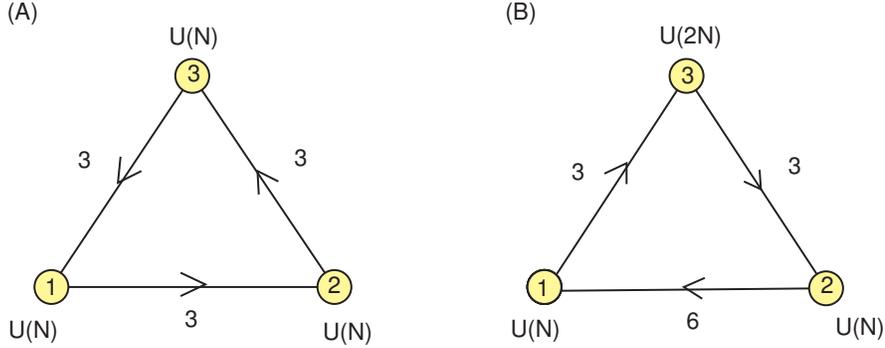}
               }
\end{center}
\vspace{-.5cm} \caption{ \it (A) Quiver diagram associated to the
exceptional collection (\ref{P2standard}). (B) Dual quiver
diagram, obtained from (A) by Seiberg duality on node
(3).}\label{P2quiver}
 \end{figure}

For ${\bf P}^2$ we will take the customary exceptional collection
\be\label{P2standard} 1.\  \cO(0) \qquad 2.\ T(-1) \qquad  3.\
\cO(1) \ee
The maps are given by
\be X_{12} = A^i {\del_i} \qquad X_{23} = \langle \bullet, B^j
{\del_j}\rangle \qquad X_{13} =  C_k z^k \ee
Here we have written the NC deformation of the identification
$\Lambda^2 T_X\otimes \cO(2)  \sim \cO(1)$ as $\langle \del_{i},
\del_{j} \rangle = g_{ijk} z^k$ for some tensor $g_{ijk}$. In the
commutative case, $f_{ijk} = \epsilon_{ijk}$, but in the
non-commutative one needs  some care in defining the bundles and
this relation will be continuously deformed. Since $z^i \del_i$ is
a trivial tangent vector, we have $g_{ijk} z^j z^k =g_{ijk} z^i
z^k=0$, hence $g_{ijk} = f_{ijk}$.

From the composition of maps one finds the expected superpotential
\be
W = f_{ijk} A^i B^j C^k .
\ee
We can rewrite this as
\ba
W &=& \lambda_1 \epsilon_{ijk}A^iB^jC^k + \lambda_2 s_{ijk}A^iB^jC^k +
\lambda_3 (A^1 B^1 C^1 + A^2 B^2 C^2 + A^3 B^3 C^3) \eol
\ea
where $s_{ijk} = |\epsilon_{ijk}|$ is a symmetric tensor. As
mentioned before these are just the Leigh-Strassler deformations
of $N=4$ Yang-Mills orbifolded by $Z_3$. A deformation by
$\lambda_2$ is called the $\beta$-deformation
\cite{Benvenuti:2005wi,Lunin:2005jy}.

To find the moduli space we should solve the F- and D-term
equations. Let us just consider the case of a single D3-brane. We
can be brief because the F-term equations for $C^k$ were already
discussed in the previous subsection. The result of that discussion
was that for generic values of the NC parameters the set of
solutions is just the embedded commutative curve in ${\bf P}^2$. If
we also consider VEVs for $C^k$ then we can also move the D3-brane
in the radial direction and the moduli space is just the cone over
the elliptic curve. For a larger number of D3-branes one obtains a
much more interesting structure however, for instance for special
discrete values of the NC parameters new branches seem to open up
where the branes form some fuzzy geometry \cite{Myers:1999ps}. Such
structure should appear when the automorphism $\sigma$ is of finite
order. In the context of mass deformations of $N=4$ Yang-Mills
theory ($N=1^*$) this was first investigated in [Polch-Strass], and
in the context of marginal deformations this was investigated in
\cite{Berenstein:2000te,Dorey:2004xm,Benini:2004nn}.

The fact that the superpotential is built on $f_{ijk}$ is actually
not too surprising. Even though the $PGl(3,{\bf C})$ symmetry of
the ${\bf P}^2$ is broken, there is still a quantum group symmetry
that uniquely fixes the superpotential. The tensor $f_{ijk}$
corresponds to the quantum determinant ${\bf 3} \otimes {\bf 3}
\otimes {\bf 3} \to {\bf C}$.

\newsubsection{The projective plane, revisited}

If we take the exceptional collection $\{ \cO(0),T(-1),\cO(1) \}$
and move $T(-1)$ one spot to the right, we get the exceptional
collection%
\be 1. \  \cO(0) \qquad 3.\ \cO(1) \quad 2.\ \cO(2) \ee
where $\cO(2) = R_{\cO(1)}T(-1)$. Let us try to understand the
quiver directly from this collection.

In order to describe a D3 brane, we consider the resolution
\be
\cO(0)[-2] \to \cO(1)^2[-1] \to \cO(2)[0] \to \cO_p
\ee
Taking into account the shifts in the derived category (spectral flow), we get
\ba \Hom(\cO(0),\cO(1)) \ &\to&
%\Ext^1(i_*\cO(0)[-2],i_*\cO(1)[-1])
N_{\rm gh} = 0 - (-2) + (-1) = +1 \eol
 \Hom(\cO(1),\cO(2)) \ &\to & %\Ext^1(i_* \cO(1)[-1],i_* \cO(2)[0])
N_{\rm gh} = 0 - (-1) + 0 = +1 \eol
 \Hom(\cO(0),\cO(2))^* &\to &
N_{\rm gh} = 3 - (0 - (-2) + 0) = +1
%\Ext^1(i_* \cO(2)[0],i_*\cO(0)[-2])  .%\eol
\ea
Therefore the bifundamentals all come from BRST cohomology classes
at ghost number one, as required. The associated quiver diagram is
drawn in figure \ref{P2quiver}B. We have the maps
\be\label{P2exc} X_{13} = A_i z^i  \qquad X_{32} = B_i z^i \qquad
X_{12} = C_{ij} z^i z^j \ee
Here we defined $C_{ij}^* = C^{ij}$ to have nine components,
whereas $\cO(2H)$ has only six generators. We can account for the
difference by adding three Lagrange multiplier fields
$Z_1,Z_2,Z_3$ and adding the following mass terms to the
superpotential:
\be W_{\rm mass}= f_{ijk}C^{ij} Z^k. \ee
Then we have the following non-commutative generalisation of the usual superpotential:
\be
W = A_i B_j C^{ij} + f_{ijk} C^{ij} Z^k.
\ee
If desired one can explicitly integrate out massive fields. If we solve for $C^{21},C^{31}$ and $C^{32}$, we obtain
\ba W &=& (\beta A_1 B_1 - \gamma A_3 B_2) C^{11} + (\beta A_1 B_2
- \alpha A_2 B_1) C^{12}
 +(-\alpha A_1 B_3 + \beta A_3 B_1) C^{13} \eol
 & & + (\beta A_2 B_2 - \gamma A_1 B_3) C^{22} +(\beta A_2 B_3 - \alpha A_3 B_2) C^{23}
  + (\beta A_3 B_3 - \gamma A_2 B_1) C^{33} .\eol
\ea
If there is a quantum group symmetry, the superpotential is again the unique one obtained from
picking the singlet in the tensor product of representations ${\bf 3} \otimes {\bf 3} \otimes {\bf 6}$.

Let us briefly check that this result agrees with the previous
section. If we perform a Seiberg duality on node 3 we should
reproduce the $Z_3$ symmetric quiver. Thus we replace $A_i B_j$ by
the meson fields $M_{ij}$, add the dual quarks
$\tilde{A}^i,\tilde{B}^j$, and modify the superpotential:
\be
W_{\rm dual} = M_{ij} C^{ij} + f_{ijk} C^{ij} Z^k + \tilde{B}^i \tilde{A}^j M_{ij} .
\ee
After integrating out $M_{ij},C^{ij}$, we obtain
\be
W_{\rm dual} = - f_{ijk} \tilde{B}^i \tilde{A}^j Z^k
\ee
which is, up to some simple field redefinitions, identical to the superpotential we obtained earlier.

\newsubsection{${\bf P}^1 \times {\bf P}^1$}

%%%%
 \begin{figure}[th]
\begin{center}
        %\resizebox{\textwidth}{!}{
            \scalebox{0.7}{
               \includegraphics[width=\textwidth]{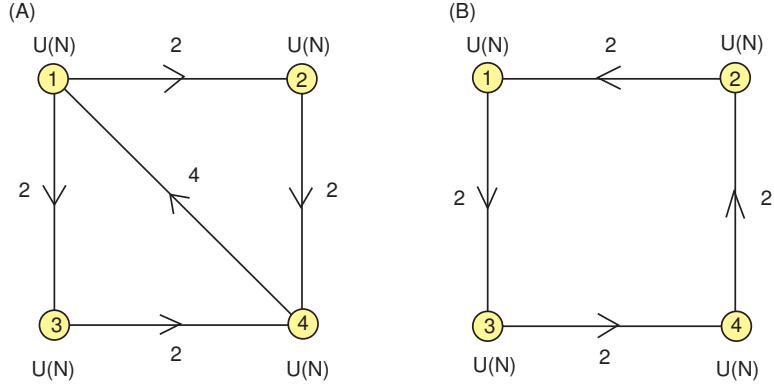}
               }
\end{center}
\label{mutationrule} \vspace{-.5cm} \caption{ \it (A): Quiver
associated with the collection (\ref{P1xP1exc}). (B): Quiver
obtained from (A) by Seiberg duality on node
2.}\label{P1xP1quiver}
 \end{figure}

We take the customary collection
\be
\begin{array}{lll}\label{P1xP1exc}
1.\ \cO(0,0)\qquad & 2.\  \cO(0,1) \qquad&  4.\ \cO(1,1) \eol
                    & 3.\ \cO(1,0) &
\end{array}
\ee
The quiver diagram is drawn in figure \ref{P1xP1quiver}. A look at
the standard quiver diagram reveals a three-dimensional space of
marginal deformations of the superpotential modulo field
redefinitions. This agrees with the geometry: there is a
9-dimensional space of Poisson deformations. Subtracting the
6-dimensional space of coordinate redefinitions leaves three
parameters.

Constructing the superpotential is relatively easy. The discussion
closely mirrors the case of ${\bf P}^2$. Let us denote the
coordinates on the ``left'' ${\bf P}^1$ by $z^\alpha$ and the
coordinates on the ``right'' ${\bf P}^1$ by $w^{\dot{\beta}}$.
Then we may define a non-commutative structure through the
equations
\ba\label{NCquadric}
0 &=& w^1 z^1 + \alpha z^1 w^1 + \delta z^2 w^2 \eol
0 &=& w^2 z^1 + \beta z^1 w^2 + \gamma z^2 w^1  \eol
0 &=& w^1 z^2 + \beta z^2 w^1 + \gamma z^1 w^2  \eol
0 &=& w^2 z^2 + \delta z^1 w^1 + \alpha z^2 w^2 .
\ea
We can write this as
\be
\left(%
\begin{array}{cccc}
  w^1 & 0 & \alpha z^1 & \delta z^2 \\
  w^2 & 0 & \gamma z^2 & \beta z^1 \\
  0 & w^1 & \beta z^2 & \gamma z^1 \\
  0 & w^2 & \delta z^1 & \alpha z^2 \\
\end{array}%
\right)
\cdot
\left(%
\begin{array}{c}
  z^1 \\
  z^2 \\
  w^1 \\
  w^2 \\
\end{array}%
\right) = 0.
\ee
The determinant of the matrix is an equation of bidegree $(2,2)$
which is an elliptic curve in ${\bf P}^1 \times {\bf P}^1$. This
is the embedded commutative curve where the Poisson structure
degenerates. For every point on this curve, the matrix has a
unique eigenvector, which determines a point in ${\bf P}^1 \times
{\bf P}^1$. The set of points obtained this way also forms an
elliptic curve, and the correspondence point$\to$ eigenvector
again yields an automorphism of the elliptic curve which we denote
by $\sigma$.

Now we use this to calculate the superpotential. The Ext generators are given by
\be
\begin{array}{rclrcl}
X_{13} &=& A_\alpha z^\alpha \qquad \qquad &X_{12} &=& C_{\dot{\alpha}} w^{\dot{\alpha}}  \eol
X_{34} &=& B_{\dot{\alpha}} w^{\dot{\alpha}} &X_{24} &=& D_\alpha z^\alpha  \eol
X_{14} &=& E_{\alpha\dot{\beta}}z^\alpha w^{\dot{\beta}} & & &
\end{array}
\ee
The superpotential is then
\ba
W &=& (C_1 D_1 + \alpha A_1 B_1 + \delta A_2 B_2)E^{11} + (C_2 D_1 + \beta A_1 B_2 + \gamma A_2 B_1) E^{12} \eol
& & + (C_1 D_2 + \beta A_2 B_1 + \gamma A_1 B_2) E^{21} + (C_2 D_2 + \delta A_1 B_1 + \alpha A_2 B_2) E^{22}.\eol
\ea
It is clear that the NC relations (\ref{NCquadric}) translate
directly into superpotential terms. The discussion above therefore
implies that the moduli space is simply the (cone over the)
embedded commutative elliptic curve.

Before closing this section let us discuss the quiver one obtains from a Seiberg duality on node 2. The quiver is drawn
in figure \ref{P1xP1quiver}B and the superpotential is given by
\ba
W_{\rm dual} \is \lambda_1 (A_1 B_1 C_2 D_2 - A_1 B_2 C_2 D_1 + A_2 B_2 C_1 D_1 - A_2 B_1 C_1 D_2) \eol
& & +\lambda_2(A_1 B_1 C_2 D_2 + A_1 B_2 C_2 D_1 - A_1 B_2 C_1 D_2 - A_2 B_1 C_2 D_1 \eol
& & \qquad + A_2 B_2 C_1 D_1 + A_2 B_1 C_1 D_2
- A_1 B_2 C_1 D_2 - A_2 B_1 C_2 D_1) \eol
& & + \lambda_3 (A_1 B_1 C_2 D_2 + A_1 B_2 C_2 D_1 + A_2 B_2 C_1 D_1 + A_2 B_1 C_1 D_2) \eol
& & +
\lambda_4(A_1 B_1 C_1 D_1 + A_2 B_2 C_2 D_2) \eol
\ea
with
\be
\begin{array}{rclrcl}
\alpha &=&   - \lambda_1 + \lambda_2 + \lambda_3 & \qquad \gamma &=& -2\lambda_2  \eol
\beta &=& \lambda_1 + \lambda_2 + \lambda_3 & \delta &=& \lambda_4 .\eol
\end{array}
\ee
This quiver is related to the conifold singularity by a $Z_2$ orbifold. Thus for our next example we turn to the conifold.

The NC deformations break the $PGl(2,{\bf C}) \times PGl(2,{\bf
C})$ symmetry of the complex structure. However at least for a
subset of the NC parameters there should still be a quantum group
symmetry.

\newsubsection{The conifold}

%%%%
 \begin{figure}[th]
\begin{center}
        %\resizebox{\textwidth}{!}{
            \scalebox{0.5}{
               \includegraphics[width=\textwidth]{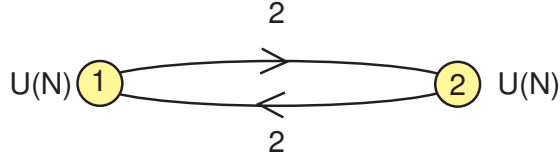}
               }
\end{center}
\label{mutationrule}
\vspace{-.5cm}
\caption{ \it The well-known conifold quiver, a $Z_2$ quotient of figure \ref{P1xP1quiver}B.}\label{conifoldquiver}
 \end{figure}

The surface ${\bf P}^1 \times {\bf P}^1$ can be embedded in ${\bf
P}^3$ through the Segre embedding. Namely if we define
$x^{\alpha\dot{\beta}} = z^\alpha w^{\dot{\beta}}$ then the image
of ${\bf P}^1 \times {\bf P}^1$ is given by the quadric surface
$x^{11} x^{22} - x^{12} x^{21} = 0 \in {\bf P}^3$. If we regard
this as an equation in affine 4-space then we do not get the cone
over ${\bf P}^1 \times {\bf P}^1$ but a double cover of it. This
is of course the well-known conifold singularity. To recover the
cone over ${\bf P}^1 \times {\bf P}^1$, we have to perform a $Z_2$
orbifold of the conifold, given by $x^{\alpha\dot{\beta}}\to
-x^{\alpha\dot{\beta}}$. We can use the $Z_2$ orbifolding to
obtain the quiver diagram \ref{P1xP1quiver}B from the conifold
quiver, or conversely we can recover the conifold quiver from
\ref{P1xP1quiver}B by modding out by the $Z_2$ quantum symmetry,
which identifies the fields $A_i = C_i$ and $B_j = D_j$. The
resulting quiver is drawn in figure \ref{conifoldquiver}.

The space of marginal deformations of the superpotential has already been examined \cite{Benvenuti:2005wi}, and it was found
that there exists a 3-parameter family of deformations, just as we found for the quadric. In fact, we can use the fact that the
quivers are related by a $Z_2$ quotient to map the deformations into each other. Thus we get the following superpotential
for the conifold quiver:
\ba\label{conifoldpotential}
W_{\rm conifold} &=& 2 \lambda_1 (A_1 B_1 A_2 B_2 - A_1 B_2 A_2 B_1) \eol
& & + 2\lambda_2(A_1 B_1 A_2 B_2 + A_1 B_2 A_2 B_1 - A_1 B_2 A_1 B_2 - A_2 B_1 A_2 B_1) \eol
& & + 2 \lambda_3 (A_1 B_1 A_2 B_2 + A_1 B_2 A_2 B_1) \eol
& & +
\lambda_4(A_1 B_1 A_1 B_1 + A_2 B_2 A_2 B_2) %\eol
\ea
%
%By quotienting a discrete subgroup, one finds the 3 NC deformations of ${\bf P}^1 \times {\bf P}^1$ and
%Del Pezzo 7 (at a special complex structure point).

The same idea can now be used to obtain the NC structures on the
conifold. Again we define $x^{\alpha\dot{\beta}} = z^\alpha
w^{\dot{\beta}}$ except that $z^\alpha, w^{\dot{\beta}}$ no longer
commute but instead satisfy (\ref{NCquadric}). This will lead to a
deformation of the seven equations $x^{11} x^{22} - x^{12} x^{21}
= 0$ and $x^{\alpha\dot{\beta}} x^{\gamma\dot{\delta}} -
x^{\gamma\dot{\delta}} x^{\alpha\dot{\beta}}=0$. Using a Gr\"obner
basis computation we find the following relations:
\ba 0 &=& \alpha\left( \gamma ^2-  \beta ^2\right) x^{2\dot{1}}
x^{1\dot{2}}+\gamma\left(
   \delta ^2-\alpha ^2 \right) x^{2\dot{1}} x^{2\dot{1}}+\beta\left(\alpha ^2 -  \delta^2
   \right) x^{2\dot{2}} x^{1\dot{1}}\eol
   & & \qquad +\delta\left(\beta ^2  -\gamma ^2 \right)
   x^{2\dot{2}} x^{2\dot{2}} \eol
0 &=& \delta \left(\beta ^2  -\gamma ^2  \right)
   x^{2\dot{1}} x^{1\dot{1}}+\beta\left(\alpha ^2 - \delta ^2\right)
   x^{2\dot{1}} x^{2\dot{2}}+\gamma\left(  \delta ^2-\alpha ^2  \right)
   x^{2\dot{2}} x^{1\dot{2}}\eol
   & & \qquad +\alpha\left(  \gamma ^2-  \beta ^2\right)
   x^{2\dot{2}} x^{2\dot{1}}\eol
0   &=&-\delta  x^{1\dot{2}} x^{1\dot{1}}+\beta  x^{1\dot{2}}
x^{2\dot{2}}-\alpha
   x^{2\dot{2}} x^{1\dot{2}}+\gamma  x^{2\dot{2}} x^{2\dot{1}}\eol
0 &=& -\beta  x^{1\dot{1}} x^{2\dot{2}}+\gamma
   x^{1\dot{2}} x^{1\dot{2}}-\gamma  x^{2\dot{1}} x^{2\dot{1}}+\beta  x^{2\dot{2}} x^{1\dot{1}}\eol
0 &=&-\delta
   x^{1\dot{1}} x^{2\dot{1}}+\gamma  x^{1\dot{2}} x^{2\dot{2}}-\alpha  x^{2\dot{1}} x^{2\dot{2}}+\beta
   x^{2\dot{2}} x^{2\dot{1}}\eol
0 &=& \delta\left(\gamma ^2  -\beta ^2  \right)
   x^{1\dot{1}} x^{1\dot{2}}+\alpha\left(  \beta ^2-  \gamma ^2\right)
   x^{1\dot{2}} x^{2\dot{2}}+\gamma\left(\alpha ^2  - \delta ^2\right)
   x^{2\dot{1}} x^{2\dot{2}}\eol
   & & \qquad +\beta\left(  \delta ^2-\alpha ^2  \right)
   x^{2\dot{2}} x^{1\dot{2}}\eol
0 &=& \delta\left(\beta ^2  -\gamma ^2  \right)
   x^{1\dot{1}} x^{1\dot{1}}+\alpha \left(  \gamma ^2-  \beta ^2\right)
   x^{1\dot{2}} x^{2\dot{1}}+\gamma\left(  \delta ^2-\alpha ^2  \right)
   x^{2\dot{1}} x^{2\dot{1}}\eol
   & & \qquad +\beta\left(\alpha ^2  -  \delta ^2\right)
   x^{2\dot{2}} x^{1\dot{1}}
\ea
%
%Alternatively one could define $x_{\alpha\dot{\beta}} = A_\alpha B_{\dot{\beta}}$ and use the F-term equations
%from the superpotential (\ref{conifoldpotential}), which should yield the same relations.
%WRONG: the moduli space of point modules is strictly smaller
Alternatively one could start with a non-commutative structure on ${\bf C}^4$, perform an NC small resolution
of the conifold, and use the method of \cite{Aspinwall:2004bs} to derive the superpotential. This is algebraically
more complicated, so we chose to exploit the relation to ${\bf P}^1 \times {\bf P}^1$.

For the quadric (and hence, through our earlier remark, for the conifold) mathematicians have
developed the
following picture \cite{1999math.....10082S}: we start with the
4-dimensional Sklyanin algebra, which defines a non-commutative structure on ${\bf C}^4$:
\begin{equation}
\begin{array}{rcl}
x^0 x^1 - x^1 x^0 &=& \alpha_1 (x^2 x^3 + x^3 x^2)  \eol
x^0 x^2 - x^2 x^0 &=& \alpha_2 (x^3 x^1 + x^1 x^3)  \eol
x^0 x^3 - x^3 x^0 &=& \alpha_3 (x^1 x^2 + x^2 x^1)  \eol
x^2 x^3 - x^3 x^2 &=& x^0 x^1 + x^1 x^0  \eol
x^3 x^1 - x^1 x^3 &=& x^0 x^2 + x^2 x^0  \eol
x^1 x^2 - x^2 x^1 &=& x^0 x^3 + x^3 x^0  \eol
\end{array}
\end{equation}
where the $\alpha_i$ are parameters satisfying $\alpha_1 +
\alpha_2 + \alpha_3 + \alpha_1\alpha_2\alpha_3=0$. The center of
this algebra is generated by two quadratic Casimir elements
\be
C_1 =  x_0^2 + J_1 x_1^2 + J_2 x_2^2 + J_3 x_3^2, \qquad \qquad C_2 = x_1^2 + x_2^2 + x_3^2
\ee
The $J_i$ can be determined in terms of the $\alpha_i$. This
defines a three parameter family of NC structures on the conifold
$C_1+ \lambda C_2=0$. If desired, one can do a coordinate
transformation so that the conifold is written in the standard
form and all the parameters appear in the Sklyanin algebra. To get
the NC structures on the cone over the quadric we should simply
quotient by $x^i \to -x^i$. The locus $C_1 = C_2 = 0$ is the
embedded commutative locus, a cone over the elliptic curve in the
quadric. If we put a single D3-brane at the singularity, then this
commutative locus is generically the moduli space of the gauge
theory. Presumably $\{ x^i,\alpha_i, \lambda \}$ and our variables
$\{ x^{\alpha{\dot{\beta}}}, \alpha, \beta, \gamma, \delta\}$ are
related through coordinate redefinitions.

It is also interesting to consider the non-commutative analogue of
the conifold transition
\cite{Klebanov:2000hb,Maldacena:2000yy,Vafa:2000wi}. To this end
one puts $M$ fractional D3-branes and one ordinary D3-brane at the
conifold. This yields the same quiver theory except that the gauge
group is $U(M+1) \times U(1)$. In the IR this is effectively an
$SU(M+1)$ gauge theory with two quarks and two anti-quarks.
Therefore we expect that the Affleck-Dine-Seiberg superpotential
gets generated and our total superpotential is
\be\label{defconifold} W_{\rm total} = W_{\rm conifold} + (M-1)
\left( {2\Lambda^{3M+1}\over {\rm det}(A_\alpha B_{\dot{\beta}}) }
\right)^{1/M-1} \ee
Now how can we find the deformation of the equations that define
the NC conifold? Note that the NC conifold is not (a component) of
the moduli space of this theory, since the D3-brane can only move
on the locus where the NC structure degenerates. On the other
hand, it is not hard to guess what it must be. To get a consistent
equation, we can only deform $C_1 +  \lambda C_2 =0$ by adding
other Casimirs of the Sklyanin algebra.\footnote{We expect that
the Sklyanin algebra itself cannot be deformed by non-perturbative
corrections, however we have not proven this statement.} Moreover,
instanton corrections come with a positive power of $\Lambda$, so
by dimension counting it must multiply a Casimir of degree less
than two (the couplings $\lambda_i$ are dimensionless). Then the
deformation should be of the form
\be C_1 + \lambda C_2 = a (\Lambda^{3M+1})^{1/M} {\bf 1} \ee
The power of $\Lambda$ is the same as in \cite{Klebanov:2000hb}.
Since the coefficient $a$ is non-zero in the commutative limit, it
should be non-zero in the non-commutative case also. Note that all
the equations are invariant under $x^i \to - x^i$, so we also
expect a transition when we put fractional branes at the
non-commutative collapsed ${\bf P}^1 \times {\bf P}^1$
singularity.

\newsubsection{Blow-ups of ${\bf P}^2$}

One can only blow-up commutative points
\cite{1999math.....10082S}, i.e the points must lie on the
elliptic curve where the NC structure degenerates. We will discuss
a three-block exceptional collection on  Del Pezzo 3 as our main
representative of the higher Del Pezzos. As was shown in
\cite{Wijnholt:2002qz} the calculations up to Del Pezzo 6 are all
extremely similar to this case.

%%%%
 \begin{figure}[th]
\begin{center}
        %\resizebox{\textwidth}{!}{
            \scalebox{0.4}{
               \includegraphics[width=\textwidth]{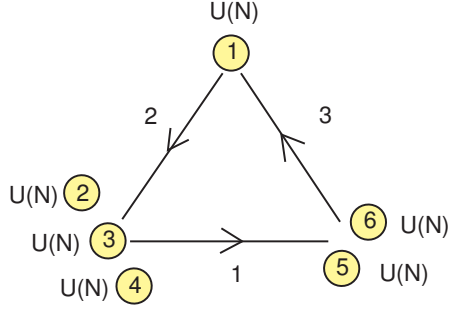}
               }
\end{center}
\vspace{-.5cm}
\caption{ \it Quiver for Del Pezzo 3 associated to the exceptional collection (\ref{dpthreeexc}).}
 \end{figure}
A simple three-block exceptional collection of line bundles is given by
\begin{equation}\label{dpthreeexc}
{ \begin{array}{lll} 1.\ \cO \qquad & 2.\ \cO(H-E_1)                 & 5.\ \cO(H)  \\
 & 3.\ \cO(H  - E_2 )\qquad  & 6.\ \cO(2H - E_1 - E_2 -E_3)  \\
  & 4.\    \cO(H-E_3)                     &    \end{array}}
\end{equation}
The exceptional curves $E_1,E_2$ and $E_3$ are obtained by blowing up the points $p,q$ and $r$.
A basis for the linear sections can be constructed as follows:
\begin{equation}
\begin{array}{rclrclrcl}
X_{12} &=& A_i z^i & \qquad X_{25} &=& 1 & \qquad X_{26} &=& \epsilon_{ijk} (q^\sigma)^i (r^\sigma)^j z^k  \eol
X_{13} &=& B_i z^i & X_{35} &=& 1 & X_{36} &=&  \epsilon_{ijk} (r^\sigma)^i (p^\sigma)^j z^k  \eol
X_{14} &=& C_i z^i & X_{45} &=& 1 & X_{46} &=& \epsilon_{ijk} (p^\sigma)^i (q^\sigma)^j z^k \eol
X_{15} &=& D_i z^i &        & &           & &                                              \eol
\end{array}
\end{equation}
Note that for $X_{12},X_{13},X_{14}$ we also added a generator
which does not vanish at $p,q,r$ respectively. We can kill these
generators by adding Lagrange multiplier fields $V_1,V_2,V_3$ and
mass terms
\be
 p^i A_i V_1 + q^i B_i V_2 + r^i C_i V_3
\ee
to the superpotential. We could of course work directly with the
massless generators, but the reason for doing it this way is that
we can write the superpotential in a much more symmetric form.

Finally we need the quadratic generators $X_{16}$, which are of
course more tricky. Sections of $\cO(2H-E_1-E_2-E_3)$ are of the
form $a_{ij} z^i z^j$, subject to the three conditions
\be
a_{ij} p^i (p^\sigma)^j =0, \qquad a_{ij} q^i (q^\sigma)^j =0, \qquad a_{ij} r^i (r^\sigma)^j =0.
\ee
A simple way to proceed is as follows. First we add the
additional sections of $\cO(2H)$ that do not vanish at $p,q,r$. We
introduce the following nine quadratic sections
\be
X_{16} =  E_{ij} z^i z^j .
\ee
and add three Lagrange multipliers $Z^1,Z^2,Z^3$ and a mass term
$f_{ijk} E^{ij} Z^k$ to get six massless fields. Then we introduce
3 additional fields $Y_1,Y_2,Y_3$ and add more mass terms to kill
the sections that do not vanish at $p,q,r$. So in total we have
\be
W_{\rm mass} = p^i A_i V_1 + q^i B_i V_2 + r^i C_i V_3 + f_{ijk} E^{ij} Z^k
%\eol & &
+ \bar{p}_i \bar{p}^\sigma_j E^{ij} Y_1
+ \bar{q}_i \bar{q}^\sigma_j E^{ij} Y_2 +\bar{r}_i \bar{r}^\sigma_j E^{ij} Y_3.
\ee
Now it is straightforward to find the following superpotential:
\ba\label{dp3pot}
W &=& W_{\rm mass} + A_{12i} X_{25} D_{51}^i + B_{13i} X_{35} D_{51}^i + C_{14i} X_{45} D_{51}^i \eol
   & & + \epsilon_{ijk}(q^\sigma)^i (r^\sigma)^j A_{12m} X_{26} E_{61}^{mk}
    + \epsilon_{ijk}(r^\sigma)^i (p^\sigma)^j B_{13m} X_{36} E_{61}^{mk}\eol
& &    + \epsilon_{ijk}(p^\sigma)^i (q^\sigma)^j C_{14m} X_{46} E_{61}^{mk} %\eol
\ea
In the commutative case we should set $f_{ijk} =\epsilon_{ijk}$,
set the automorphism $\sigma$ equal to the identity and integrate
out the massive fields. In this case one reproduces calculations
previously performed in \cite{Wijnholt:2002qz}, which are known to
yield the expected superpotential.

By turning on an expectation value for $X_{26},X_{36}$ or $X_{46}$
we get quiver theories for Del Pezzos with fewer blow-ups.

\newsubsection{Abelian orbifolds}

Consider the orbifold ${\bf C}^3/Z_k$ where the coordinates of
${\bf C}^3$ are taken to have weights $(w_1,w_2,w_3)$ under the
action of $Z_k$ (with $w_1+w_2+w_3=k$). In order to derive the
quiver gauge theory the simplest method is of course to use the
projection methods of \cite{Douglas:1996sw}. This is more powerful
than the large volume description since we also get information
about the D-terms. Nevertheless it will be useful to consider the
large volume limit. Non-commutative deformations can be described
in this framework, and it provides some insights that should apply
more generally to toric singularities and their deformations. For
recent progress in the toric case see
\cite{Franco:2005rj,Hanany:2005ss,Feng:2005gw}.

For $k > 3$ the orbifold ${\bf C}^3/Z_k$ contains  multiple
vanishing 4-cycles and we need multiple blow-ups in order to
completely resolve the singularity. After a single blow-up we get
a finite size ${\bf P}^2_{(w_1,w_2,w_3)}$ which typically has
orbifold singularities, and further blow-ups are needed to remove
these singularities. Nevertheless the weighted projective space
${\bf P}^2_{(w_1,w_2,w_3)}$ already has nice sets of exceptional
collections that we can use to construct the quiver gauge theory,
as we will now review \cite{Mayr:2000as,Wijnholt:2002qz}.

There are two canonical exceptional collections that are dual to
each other. The first is a collection of invertible sheaves
$\{R_1, \ldots, R_k \} = \{ \cO(0), \ldots, \cO(k) \}$ which is
called the bosonic basis. The non-zero cohomology groups are
$\Hom(R_i,R_j)$ which is generated by the polynomials of total
degree $j-i$ in the coordinates $z^i$. The compositions of these
maps are the obvious ones. The number of generators can be read
off from the coefficient of $h^{j-i}$ of the bosonic generating
function (the Hilbert series of ${\bf P}^2_{(w_1,w_2,w_3)}$)
\be\label{hilbert}
\chi = (1 - h^{w_1})^{-1} (1-h^{w_2})^{-1} (1-h^{w_3})^{-1} .
\ee
Although this exceptional collection is very simple it does not
lead to physical quiver diagrams for $k>3$. One could in principle
use mutations to get a physical collection as explained in
section. However it is easier to use the other canonical basis
which leads directly to the expected orbifold quiver.

The second collection is called the fermionic basis $\{ S_1,
\ldots, S_k \}$. The exact definition of the $S_i$ is a little
murkier but they are roughly of the form $\Lambda^m T \otimes
\cO(n)$. However it is easy to say what the cohomology groups are:
the non-zero ones are $\Hom(S_i, S_j)$ which is generated by
contractions with tangent vectors $\imath_{\del_i}$ of total
degree $-(j-i)$.\footnote{This is dual to wedging with the
differentials $dz^i$} The number of generators can be read of from
a fermionic generating function which is just the  inverse of
(\ref{hilbert}):
\be \chi^{-1} = (1 - h^{w_1}) (1-h^{w_2}) (1-h^{w_3}) . \ee
The fermionic basis can be obtained from the bosonic basis (up to
tensoring by an invertible sheaf) by the mutations $\{ S_1,
\ldots, S_k \} = \{ L^{k-1} R_k, L^{k-2} R_{k-1}, \ldots, R_1 \}$.
The collections are dual in the sense that $\chi(R_i,S_j) =
\delta_{ij}$.

For generic $(w_1,w_2,w_3)$ the orbifold ${\bf C}^3/Z_k$ admits
only one NC deformation:
\be
xy = qyx, \qquad yz = q zy, \qquad zx = q xz % \qquad {\rm with } \ \ q^k = e^{it}
\ee
The commutatation relations of $\imath_{\del_i}$ can be deduced
for instance from the fact that the fermionic basis is dual to the bosonic basis \cite{2004math......4281A}:
\be
\imath_{\del_x} \imath_{\del_y} = -q\, \imath_{\del_y} \imath_{\del_x},
\quad \imath_{\del_y} \imath_{\del_z}  = -q\, \imath_{\del_z} \imath_{\del_y},
\qquad \imath_{\del_z} \imath_{\del_x}  = -q\,  \imath_{\del_x} \imath_{\del_z}.
\ee
Using these relations, one finds a deformation of the orbifold theory. It is the same as the
$\beta$-deformation.
%%%
 \begin{figure}[th]
\begin{center}
        %\resizebox{\textwidth}{!}{
            \scalebox{0.4}{
               \includegraphics[width=\textwidth]{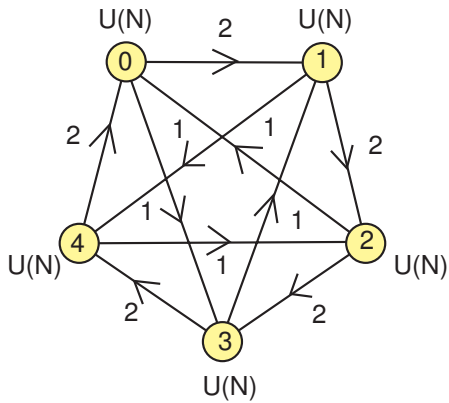}
               }
\end{center}
\label{mutationrule} \vspace{-.5cm} \caption{ \it The $Z_5$
symmetric quiver of ${\bf C}^3/Z_5$.}\label{Z5quiver}
 \end{figure}
Let us consider as an example the orbifold ${\bf C}^3/Z_5$, the
extension to other cases being straightforward. We find the
superpotential
\ba
W \is (Y_{01} X_{12} - { q}X_{01} Y_{12} ) Z_{20} +
 (Y_{12} X_{23} - { q}X_{12} Y_{23} ) Z_{31} +
 (Y_{23} X_{34} - { q}X_{23} Y_{34} ) Z_{42} \eol
& &+
  (Y_{34} X_{40} -{ q} X_{34} Y_{40} ) Z_{03} +
 (Y_{40} X_{01} - { q}X_{40} Y_{01} ) Z_{14}
\ea
For special $(w_1,w_2,w_3)$ there may exist additional
deformations. We expect that if the non-commutative deformations
are written as $f_{ijk} z^i z^j = 0$ then the superpotential is of
the form $W = f_{ijk} X^i Y^j Z^k$, where $X,Y,Z$ are the
projected adjoint fields of the parent $N=4$ theory.

\newpage

\newsection{Quivers with ghosts and generalised Seiberg dualities}

One of the problems with simple exceptional collections is
that they typically contain ghosts. Recall that when we build
quiver diagrams out of a set of fractional branes, we must ensure
that all the bifundamental fields correspond to vertex operators
at ghost number one (in the derived category sense). If some of
the bifundamentals have the wrong ghost number, we do not seem to
be able to construct a sensible gauge theory.

Nevertheless we will show that one can consistently manipulate such
quivers at the level of F-terms. As we discussed in section 2, the
idea is to say that every cohomology class of ghost number $p<2$ on
the Calabi-Yau gives rise to a chiral field in four dimensions with
physical ghost number $p$.

The main object here is to understand how the quiver theories for
different exceptional collections are related. In order to do this
we will first discuss the quiver for a brane/anti-brane
pair\footnote{The idea of adding an anti-brane has also been
considered in \cite{Berenstein:2002fi}. However our treatment of the
open string modes will be rather different.}. Basically the massless
open strings between such a pair gives rise to a set of fields and
ghost fields which cancel each other precisely. The field/ghost
field pairs can then be used to rearrange the degrees of freedom in
a quiver to a dual quiver.

Along the way we will also get a new perspective on Seiberg duality.

\newsubsection{The brane/anti-brane quiver}

Let us first consider a brane/anti-brane pair in
isolation. Such a pair can be
regarded as a complicated description of `nothing.'\footnote{Other
examples of systems without physical excitations are the
$bc\beta\gamma$ quartets in 2-dimensional CFT.} After that we will
add such pairs of `nothing' to our quiver theories and use them to
rearrange the degrees of freedom. The rearranged quiver will be
the Seiberg dual theory of the original quiver.

Consider two identical copies of a brane, $F_1$ and $F_2$. Then
the massless spectrum is as follows: we have two ghost number zero
operators $\Ext^0(F_1,F_1)$ and $\Ext^0(F_2,F_2)$ which are just
the identity map. These correspond to the two $U(1)$ vector
multiplets for each brane. We also have a generator from
$\Ext^0(F_1,F_2)$ and another from $\Ext^0(F_2,F_1)$. These
correspond to the $W^\pm$ bosons, and altogether we therefore have
a $U(2)$ vector multiplet.

%%%%
 \begin{figure}[th]
\begin{center}
        %\resizebox{\textwidth}{!}{
            \scalebox{.3}{
               \includegraphics[width=\textwidth]{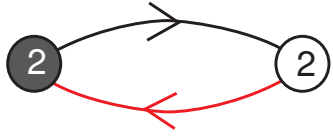}
               }
\end{center}
\vspace{-.5cm} \caption{ \it Quiver diagram for the brane/anti-brane system.}\label{bbbar}
 \end{figure}

Now we can apply one unit of spectral flow to one of the branes in
order to turn it into an anti-brane. There are basically two
choices, we can shift $F_1$ up or down with respect to $F_2$. We
will shift $F_1$ to $F_1[-1]$. The effect of this is to shift the
ghost numbers of the open strings stretching between the two
branes: the open string stretching from $F_1$ to $F_2$ will now
have ghost number $N_{\rm gh} = +1$, and the string stretching
from $F_2$ to $F_1$ will have ghost number $N_{\rm gh} = -1$. In
summary:
\ba N_{\rm gh} = 1: & & X_{\bar{2}2} \eol N_{\rm gh} = 0: & &
\Lambda_2,\ \Lambda_{\bar{2}} \eol N_{\rm gh} = -1: & &
\Upsilon_{2\bar{2}} \ea
The ghost number zero fields generate the following symmetries:
\be\label{trans1} \delta X_{\bar{2}2} = \Lambda_{\bar{2}} X_{\bar{2}2} -
X_{\bar{2}2} \Lambda_{{2}} , \qquad \delta \Upsilon_{2\bar{2}}
=\Lambda_{{2}} \Upsilon_{2\bar{2}} - \Upsilon_{2\bar{2}}
\Lambda_{\bar{2}} \ee
Moreover the ghost number minus one field generates a redundancy:
\be\label{trans2} \delta \Lambda_2 = \Lambda_{2\bar{2}} X_{\bar{2}2},
\qquad\delta \Lambda_{\bar{2}} = X_{\bar{2}2} \Lambda_{2\bar{2}}
\ee
We can write all this in a more compact form using the BV formalism
(for reviews see \cite{Witten:1990wb,Weinberg:1996kr}). We introduce
the anti-fields, $\{ X^*_{2\bar{2}}, \Lambda_2^*,\
\Lambda_{\bar{2}}^*, \Upsilon_{\bar{2}2}^*\}$, of ghost numbers
$\{2,3,3,4\}$ respectively, and the following bracket:
\be \{ A, B \} = \sum_i {\del_R A\over \del X_i} {\del_L B\over \del
X_i^*} -{\del_R A\over \del X_i^*} {\del_L B\over \del X_i}
% \sim \sum_i \del_{X_i} A \, \del_{X^*_i} B - \del_{X_i^*}  A\, \del_{X_i} B
\ee
Here $\del_R,\del_L$ denote right and left differentiation. Then we
can define an extended superpotential which is a function of all the
fields and anti-fields, such that gauge transformations are
generated by $W$ itself
\be
\delta A = \{ W, A \}.
\ee
If we pick the following superpotential:
\be\label{babsuper} W = X_{2\bar{2}}^*( \Lambda_{\bar{2}}
X_{\bar{2}2} - X_{\bar{2}2} \Lambda_{{2}}) + \Upsilon_{\bar{2}2}^*
(\Lambda_{{2}} \Upsilon_{2\bar{2}} - \Upsilon_{2\bar{2}}
\Lambda_{\bar{2}}) + \Lambda_2^* \Upsilon_{2\bar{2}} X_{\bar{2}2} +
\Lambda_{\bar{2}}^* X_{\bar{2}2} \Upsilon_{2\bar{2}} \ee
defined on the extended phase space of the B-model, then we
reproduce gauge variations (\ref{trans1}),(\ref{trans2}). Moreover
with this superpotential the BV master equation is satisfied
\be
\{ W,W \} = 0
\ee
which just says that the superpotential itself is gauge invariant.

The superpotential (\ref{babsuper}) may presumably be derived more
systematically along the following lines. We start with the quiver
for two ordinary branes, which has a $U(2)$ gauge symmetry. The
extended superpotential in this case is simply
\be\label{gaugesuper} W = \half \Lambda^{c*} \Lambda_a \Lambda_b \,
f^{ab}_c \ee
where $f^{ab}_c$ are the structure constants of $U(2)$, $\Lambda_a$
are the ghost number zero generators of the gauge symmetry (recall
they are anti-commuting), and $\Lambda^{c*}$ the corresponding
anti-ghosts. The identity $\{ W,W \} = 0$ reduces to the Jacobi
identity. Now we apply one unit of spectral flow to the second
brane. This shifts the ghost numbers of suitable linear combinations
of the $\Lambda_a$. There are some sign conventions which we have
not completely figured out, but with some suitable signs this
procedure should turn (\ref{gaugesuper}) into (\ref{babsuper}).

For the brane/anti-brane quiver, one cannot construct any gauge
invariant operators out of the ghost number one field, so the moduli
space consists just of a single point. If we turn on a VEV for the
$N_{\rm gh} = +1$ mode, all the degrees of freedom cancel pairwise,
and the only state left is the vacuum \cite{Vafa:2001qf}. The ghost
number +1 field cancels with the anti-symmetric combination
$\Lambda_2 = - \Lambda_{\bar{2}}$, and the ghost number -1 field
cancels with the symmetric combination $\Lambda_2 = +
\Lambda_{\bar{2}}$. In the extended superpotential, this is
manifested as quadratic terms for the fields after we turn on a VEV.
This is our model of `nothing.'

Shortly after this paper appeared, it was suggested that in the full
ten-dimensional string theory we should interpret topological
anti-branes not as ordinary anti-branes but as `ghost-branes.' The
worldvolume theory of $N$ branes and $M$ ghost-branes in the full
ten-dimensional string theory should be ${\cal N}=1$ SUSY Yang-Mills
theory with the supergroup $U(N|M)$ as gauge group
\cite{Okuda:2006fb}. Due to cancellations in gauge invariant
correlation functions, this would give the same answers as in
$U(N-M)$ Yang-Mills theory. This is indeed very reminiscent of the
structure we have found here. However there is still a puzzle. From
the supergroup point of view the ghost number one field
$X_{\bar{2}2}$ should be the internal part of the vertex operator
for an off-diagonal gauge field of the supergroup. It seems more
natural however to say that it gives rise to a physical chiral field
in four dimensions. As we will see this will be quite crucial for us
because $X_{\bar{2}2}$ is going to give rise to some of the magnetic
quarks of the Seiberg dual theory, which are chiral fields. It would
be interesting to elucidate this issue.

\newsubsection{The mechanism behind Seiberg duality}

%%%%
 \begin{figure}[th]
\begin{center}
        %\resizebox{\textwidth}{!}{
            \scalebox{.8}{
               \includegraphics[width=\textwidth]{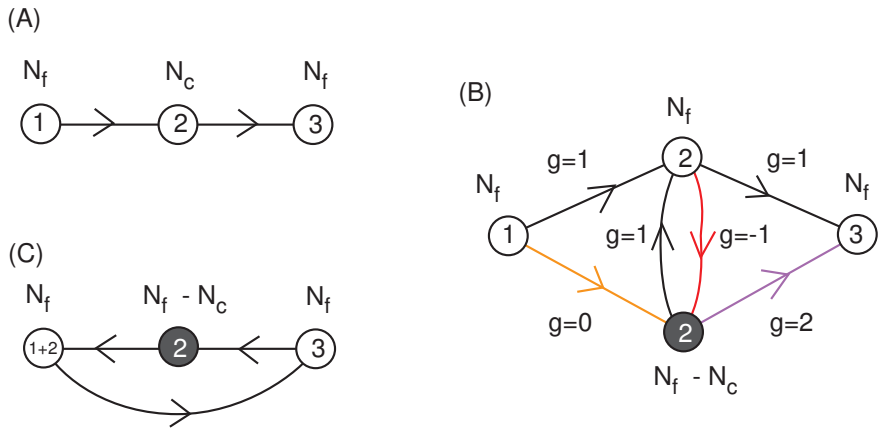}
               }
\end{center}
\vspace{-.5cm} \caption{ \it (A): Quiver for SUSY QCD with $N_c$
colours and $N_f$ flavours. (B): Quiver obtained by adding $N_f -
N_c$ brane/ghost-brane pairs to (A). (C): Seiberg dual obtained by
merging nodes 1 and 2.}\label{ancq}
 \end{figure}

Our discussion of the topological brane/anti-brane system puts us in
a position to give a proof of Seiberg duality at the level of
F-terms. Consider SUSY QCD as in figure \ref{ancq}A, and add $N_f-N_c$
brane/ghost-brane pairs. The quiver in
\ref{ancq}B has the following fields:
\ba
N_{\rm gh} = 1: & & X_{12}, X_{23}, X_{3\bar{2}}, X_{\bar{2}2}; \eol
N_{\rm gh} = 0: & &  \Lambda_2, \Lambda_{\bar{2}}, \Lambda_{1\bar{2}}; \eol
N_{\rm gh} = -1: & & \Upsilon_{2\bar{2}}.
\ea
Here we are taking nodes $1$ and $3$ to be non-dynamical, so we have not included
ghost number zero fields for them. We also introduce the anti-fields. By turning on a VEV
for $X_{\bar{2}2}$ we go back to the original quiver, and by turning
on a VEV for $X_{12}$ we go to the Seiberg dual.

As in the previous subsection, we can obtain the extended
superpotential for figure \ref{ancq}B, which turns out to be:
\ba W &=& X_{\bar{2}2} X_{23} X_{3\bar{2}} + X_{32}^*
\Upsilon_{2\bar{2}} X_{\bar{2}3}^* + X_{21}^*( - X_{12} \Lambda_2 +
\Lambda_{1\bar{2}} X_{\bar{2}2}) + X_{32}^*  \Lambda_2 X_{23} \eol &
& - X_{\bar{2}3}^* X_{3\bar{2}} \Lambda_{\bar{2}} + X_{2\bar{2}}^*
(\Lambda_{\bar{2}} X_{\bar{2}2} - X_{\bar{2}2} \Lambda_2) +
\Lambda_{\bar{2}1}^*( - \Lambda_{1\bar{2}} \Lambda_{\bar{2}} +
X_{12} \Upsilon_{2\bar{2}}) \eol & & + \Upsilon_{\bar{2}2}^*(
\Lambda_2 \Upsilon_{2\bar{2}} - \Upsilon_{2\bar{2}}
\Lambda_{\bar{2}}) + \Lambda_2^* \Upsilon_{2\bar{2}} X_{\bar{2}2} +
\Lambda_{\bar{2}}^* X_{\bar{2}2} \Upsilon_{2\bar{2}} + W_{\rm gauge}
\ea
where $W_{\rm gauge}$ generates non-abelian gauge transformations,
similar to (\ref{gaugesuper}). If we set all the anti-fields to
zero, then we are left over with the following expression:
\be
W = X_{\bar{2}2} X_{23} X_{3\bar{2}} .
\ee
This will of course descend to the Seiberg dual superpotential.

Let us do some quick counting. Suppose we want to go back to the original quiver by turning
on VEVs for $X_{\bar{2}2}$. Since $\delta X_{\bar{2}2} = \Lambda_{\bar{2}} X_{\bar{2}2} - X_{\bar{2}2} \Lambda_2$
is a matrix equation with $N_f(N_f - N_c)$ independent entries, this means there are
$N_f^2 + (N_f - N_c)^2 - N_f(N_f-N_c) = N_f^2 + N_c^2 -N_f N_c $ unbroken generators in $\Lambda_2$
and $\Lambda_{\bar{2}}$. Furthermore since $\delta \Lambda_2 = \Upsilon_{2\bar{2}} X_{\bar{2}2},
\delta \Lambda_{\bar{2}} =X_{\bar{2}2} \Upsilon_{2\bar{2}}$, the $\Upsilon_{2\bar{2}}$ pair up with
an additional $N_f(N_f-N_c)$ generators in $\Lambda_2$
and $\Lambda_{\bar{2}}$, leaving just $N_c^2$ generators, associated with node 2 in the original quiver diagram.
Furthermore, because of $\delta X_{12} = \Lambda_{1\bar{2}} X_{\bar{2}2}$, the $N_f(N_f-N_c)$ generators
in $\Lambda_{1\bar{2}}$ pair up with an equal number of the $X_{12}$, leaving us just with
the $N_f N_c$ generators in $X_{12}$ as in the original quiver diagram. Similarly, turning on
$X_{\bar{2}2}$ yields a mass term for $X_{23}$ and $X_{3\bar{2}}$, and the massless survivors
are precisely the original fields.

Instead we could turn on $X_{12}$, which takes us to the Seiberg dual. The $2N_f^2$ degrees of freedom
in $\Lambda_1, \Lambda_2$ are broken to the diagonal $N_f^2$, in the process of which the
$N_f^2$ degrees of freedom in $X_{12}$ get eaten. Also because of $\delta \Lambda_{1\bar{2}} =
X_{12} \Upsilon_{2\bar{2}}$, the $N_f(N_f-N_c)$ degrees of freedom in $\Lambda_{1\bar{2}}$ pair up
with the $N_f(N_f-N_c)$ degrees of freedom in $\Upsilon_{2\bar{2}}$. This leaves us with the Seiberg dual.

The manipulation just performed gives an equivalence at the level of
F-terms. It can clearly not be extended to the full theory because
Seiberg duality is not an exact duality. Nevertheless this gives a
new perspective on how the degrees of freedom in two dual theories
are related.

It is tempting to interpret the extended quiver \ref{ancq}B as
$U(N_f|N_f - N_c)$ SUSY gauge theory with $N_f$ quarks and $N_f$
anti-quarks. However from this supergroup point of view the magnetic
quark fields $X_{\bar{2}2}$ should correspond to off-diagonal vector
superfields of the supergroup. This point remains to be clarified.

\newsubsection{Superpotentials via quivers with ghosts}

As we reviewed in section 2, given an exceptional collection of
sheaves which generate the derived category (i.e. `fractional
branes'), one may obtain another set by applying an operation
known as a `mutation'. While the information contained in any of
the exceptional collections is equivalent, it is frequently much
easier to extract from one collection than from another. Thus one
would like a simple set of rules to obtain to transform this
information under mutation. So far such a  set of rules is known
only for exceptional collections that are related by Seiberg
duality. Here we discuss a set of rules that is meant to apply for
arbitrary mutations, which one may view as `generalised Seiberg
dualities.' Using this set of rules in principle makes the
computations of the superpotential much more systematic. For
instance for the Del Pezzo singularities we can take the
exceptional collection
\be\label{simpleexc}
\cO(0), \cO(H), \cO(2H), \cO_{E_1},\ldots, \cO_{E_n} .
\ee
to write down a quiver and superpotential. Clearly this is
essentially the same computation for all the Del Pezzo surfaces.

In the following we
will start with unphysical  but simple to understand quivers which
have bifundamental ghosts; such ghosts will be indicated with coloured
arrows. The game is then to apply mutations to get rid of the coloured
lines, and end up with a physical quiver.

In order to carry out this procedure we would like a method for
deriving the superpotential of the mutated quiver from the original
one, without having to do any new calculations with the mutated
fractional branes. We saw that for two quivers related by Seiberg
duality there is a well-defined method for writing down an
intermediate quiver and an extended superpotential by adding
brane/anti-brane pairs. We can do the same thing for quivers that
are related by a general mutation. We first illustrate the issues in
a well-known example based on ${\bf P}^1 \times {\bf P}^1$. Then we
show how it applies to exceptional collections of the form
(\ref{simpleexc}) for the Del Pezzo surfaces. Along the way, we will
see that quantities which only depend on holomorphic data, such as
the $a$-anomaly and the number of dibaryons, can be correctly
recovered from quivers with ghosts.

\newsubsection{The quadric: mutation, $a$-maximization, dibaryons}

%%%%
 \begin{figure}[th]
\begin{center}
        %\resizebox{\textwidth}{!}{
            \scalebox{.5}{
               \includegraphics[width=\textwidth]{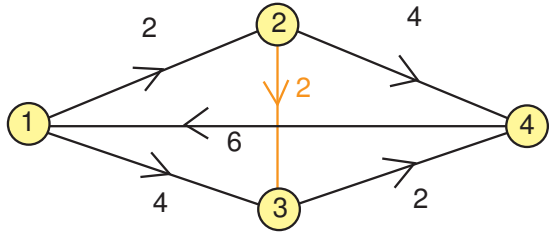}
               }
\end{center}
\vspace{-.5cm} \caption{ \it Quiver associated to the exceptional
collection (\ref{ghostcollection}). The ghost fields are indicated
by a coloured arrow.}\label{quadghostquiver}
 \end{figure}
Our favourite example of a quiver with ghosts is based on the following exceptional collection
on ${\bf P}^1 \times {\bf P}^1$ :
\be
\begin{array}{lll}\label{ghostcollection}
 1. \ \cO(-2,-1)\qquad & 2.\  \cO(-1,-1) \qquad & \ 4.\ \cO(0,0) \eol
                 & 3. \ \cO(-1,0)
\end{array}
\ee
on ${\bf P}^1 \times {\bf P}^1$. The role of the $\Ext^0$ in this
quiver was explained to us by Sheldon Katz as part of a project
\cite{oldquiver}. Similar observations since then were made
independently in \cite{GreevyHell,Aspinwall:2004vm,Herzog:2004qw}.
For simplicity, we only consider the commutative case in this
subsection.

The physical fields are given by
\be
\begin{array}{rclrcl}
X_{12} &=& A_\alpha z^\alpha  & X_{24} &=& B_{\alpha \dot{\beta}} z^\alpha w^{\dot{\beta}}  \eol
X_{13}  &=&C_{\alpha\dot{\beta}} z^\alpha w^{\dot{\beta}} & X_{34} &=& D_{\alpha} z^\alpha  \eol
X_{14} &=& E_{\alpha\gamma \dot{\beta}}z^\alpha z^\gamma w^{\dot{\beta}} \qquad &  &  &
\end{array}
\ee
After taking spectral flow into account, these correspond to vertex operators of ghost number one.
However we also have cohomology classes of ghost number zero:
\be
X_{23} = F_{\dot{\beta}}w^{\dot{\beta}}\ .
\ee
These are indicated in red in the quiver diagram. The gauge groups
are all $U(N)$.

Applying the familiar rules, we get the superpotential
\be
W = (A_{\alpha} B_{\gamma\dot{\beta}} + C_{\alpha\dot{\beta}} D_{\gamma}) E^{\alpha\gamma\dot{\beta}} .
\ee
The ghosts generate the following symmetry:
\be
\delta C_{\alpha \dot{\beta}}= -A_\alpha F_{\dot{\beta}}, \qquad
\delta B_{\alpha \dot{\beta}} = F_{\dot{\beta}} D_\alpha
\ee
which leaves the superpotential invariant. Since as we discussed this is a redundancy, then in order to get the correct
moduli space we should mod out by all the gauge groups associated to the nodes as well as the symmetries
parametrised by $F$.

%%%%
 \begin{figure}[th]
\begin{center}
        %\resizebox{\textwidth}{!}{
            \scalebox{1.0}{
               \includegraphics[width=\textwidth]{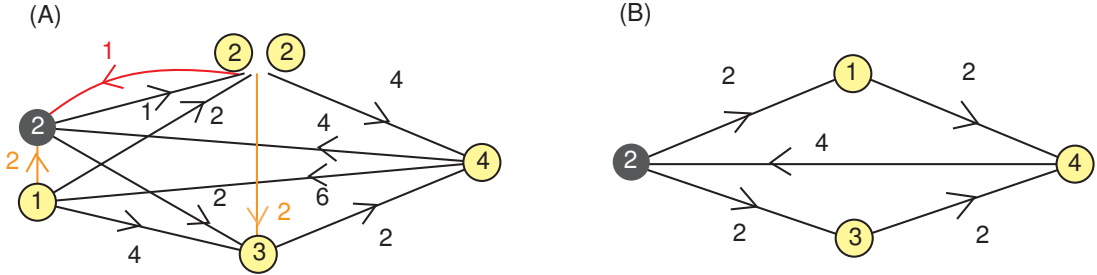}
               }
\end{center}
\vspace{-.5cm} \caption{ \it (A): Same quiver as in figure
\ref{quadghostquiver} but with brane/anti-brane pair added. (B):
Quiver obtained from (A) by condensing the links between nodes 1
and 2. This is the same quiver as in figure
\ref{P1xP1quiver}B.}\label{P1xP1Ancestor}
 \end{figure}
Now we would like to obtain a quiver without ghosts by applying a generalised Seiberg duality, i.e. a mutation.
In this case we would like to replace
\be
F_{(1)} \to R_{F_{(2)}}F_{(1)} = \cO(0,-1).
\ee
At the level of Chern characters we have
\be
{\rm ch}(\cO(0,-1) ) = -[{\rm ch}(F_{(1)}) - 2 \,{\rm ch}(F_{(2)})]
\ee
according to the Picard-Lefschetz formula. So we need two copies of $F_{(2)}$ and one copy of
$F_{(1)}$ to make $\cO(0,-1)$. We  first we do an intermediate step
by adding brane and anti-brane versions of $F_{(2)} = \cO(-1,-1)$ to get an extended
quiver diagram.
\be
\begin{array}{ccccc}
\bar{2}.\ \cO(-1,-1)[-2] &   & 2.\  \cO(-1,-1)^2[-1] &  &    \eol
\oplus   &      \to & \oplus       & \to & 4.\  \cO(0,0)[0]       \eol
1.\ \cO(-2,-1)[-2] &   & 3.\ \cO(-1,0)[-1]  &   &
\end{array}
\ee
Since the gauge group associated to node 2 has been enhanced from
$U(N)$ to $U(2N)$, there are now effectively twice as many fields
corresponding to arrows going into or out of node 2. We will label this explicitly
by introducing an index $i = 1,2$ which keeps track of which of the two nodes
with label 2 a field is connected to. In addition
we have new fields associated with the node ${\bar{2}}$:
\be
X_{\bar{2} 3} =\tilde{F}_{\dot{\beta}} w^{\dot{\beta}}, \qquad
X_{\bar{2}4} = \tilde{B}_{\alpha\dot{\beta}} z^\alpha w^{\dot{\beta}}, \qquad
X_{\bar{2}2} = U
\ee
as well as extra ghosts
\be
X_{1\bar{2}} = \tilde{A}_\alpha z^\alpha.
\ee
The quiver is drawn in figure \ref{P1xP1Ancestor}A. The new
superpotential is
\be W = (A^i_{\alpha} B^i_{\gamma\dot{\beta}} +
C_{\alpha\dot{\beta}} D_{\gamma}) E^{\alpha\gamma\dot{\beta}} +
U^i B^i_{\alpha\dot{\beta}} \tilde{B}^{\alpha\dot{\beta}} +
\tilde{F}_{\dot{\beta}} D_\alpha \tilde{B}^{\alpha\dot{\beta}}.
\ee
The symmetries are now given by
\be\label{paragauge}
\begin{array}{rclrclrcl}
\delta C_{\alpha\dot{\beta}} &=& - A^i_\alpha F^i_{\dot{\beta}}
\qquad & \delta B^i_{\alpha\dot{\beta}} &=& F^i_{\dot{\beta}}
D_\alpha \qquad & \delta\tilde{F}_{\dot{\beta}} &=& - U^i
F^i_{\dot{\beta}} \eol \delta C_{\alpha\dot{\beta}} &=&
\tilde{A}_{\alpha} \tilde{F}_{\dot{\beta}}&  \delta
\tilde{B}^{\alpha\dot{\beta}} &=& - E^{\alpha\gamma\dot{\beta}}
\tilde{A}_\gamma & \delta A^i_\alpha &=& \tilde{A}_\alpha U^i
\end{array}
\ee
The idea behind these equations is hopefully clear. For every composition
of maps we get either a superpotential term or a symmetry. When we
add the anti-branes the compositions that go through node 2 are the same
as the compositions that go through $\bar{2}$. The only possible difference
is in interpretation: when we replace 2 by $\bar{2}$, a superpotential term
may give another superpotential term or it may give a symmetry. Hence
$A^i_{\alpha} B^i_{\gamma\dot{\beta}}
 E^{\alpha\gamma\dot{\beta}} $ gives $\delta
\tilde{B}^{\alpha\dot{\beta}} = - E^{\alpha\gamma\dot{\beta}}
\tilde{A}_\gamma$.
Similarly
a symmetry may give another symmetry or it may give a superpotential term.
Hence $\delta C_{\alpha\dot{\beta}} = - A^i_\alpha F^i_{\dot{\beta}}$
gives $\delta C_{\alpha\dot{\beta}} =
\tilde{A}_{\alpha} \tilde{F}_{\dot{\beta}}$ and
$\delta B^i_{\alpha\dot{\beta}} = F^i_{\dot{\beta}}
D_\alpha $ gives $\tilde{F}_{\dot{\beta}} D_\alpha \tilde{B}^{\alpha\dot{\beta}}$.
Finally for every field that  goes through node 2 there is a new composition
involving its tilde version and the field $U$. This gives the superpotential
terms $U^i B^i_{\alpha\dot{\beta}} \tilde{B}^{\alpha\dot{\beta}}$ and the symmetries
$\delta\tilde{F}_{\dot{\beta}} = - U^i
F^i_{\dot{\beta}} ,\ \delta A^i_\alpha = \tilde{A}_\alpha U^i$.

There is one additional field which is indicated in red in the
quiver diagram. Namely apart from $X_{\bar{2}2}$ which has ghost
number one, we also have $X_{2\bar{2}}$ which has ghost number {\it
minus} one. Its action on all the fields is somewhat complicated and
is best understood by writing the superpotential as a function of
all the fields and anti-fields, as in the section on Seiberg
duality. However we will only need the action on the ghost number
zero fields, which is given by
\be \delta F^i_{\dot{\beta}} = X^i_{2\bar{2}}
\tilde{F}_{\dot{\beta}}, \qquad \delta \tilde{A}_\alpha = A^i_\alpha
X^i_{2\bar{2}} \ee

The next step is to Higgs down to the desired quiver. To this end
one turns on expectation values for all the $A$-fields. Then nodes
1 and 2, which are connected through the $A$-fields, collapse to
the single node associated to $\cO(0,-1)$. The $4N^2$ vector
multiplets that disappear became massive by eating the $4N^2$
chiral fields $A^i_\alpha$.

When we give an expectation value to $A^i_\alpha$, the two ghost
number -1 fields $X^i_{2\bar{2}}$ cancel with the two ghost number
zero fields $\tilde{A}_\alpha$. Moreover there is a mass term for
$E$ and $B$, as a result of which six of the $E$'s and six of the
$B$'s become massive and are removed from the low energy spectrum.
The remaining massless $B$'s can be parametrised by introducing
two fields $B_{\dot{\beta}}$ and setting
\be
A^i_\alpha B^i_{\gamma\dot{\beta}} = B_{\dot{\beta}}\epsilon_{\alpha\gamma}\qquad \Leftrightarrow
\qquad B^i_{\gamma\dot{\beta}} = (A^i_\alpha)^{-1} B_{\dot{\beta}}\epsilon_{\alpha\gamma}
\ee
Then after integrating out the massive degrees of freedom, we are left with the superpotential
\be
W = U^i  (A^i_\alpha)^{-1} B_{\dot{\beta}}\epsilon_{\alpha\gamma}\tilde{B}^{\gamma\dot{\beta}} +
\tilde{F}_{\dot{\beta}} D_\alpha \tilde{B}^{\alpha\dot{\beta}}.
\ee
Up to a field redefinition of the $U^i$, this is exactly the
expected superpotential for figure \ref{P1xP1Ancestor}B. Finally,
the $C$-fields are killed precisely by the symmetry generated by
$F^i_{\dot{\beta}}$. Thus we have obtained the correct quiver
theory for figure \ref{P1xP1Ancestor}B by starting with figure
\ref{quadghostquiver} and applying a generalised Seiberg duality.
This is in agreement with the idea that the $F$-term information
in any quiver obtained from an exceptional collection is
equivalent and can be related through generalised Seiberg
dualities, or mutations.

We warn the reader that the remainder of this subsection is {\it
rather formal} since we have not defined a physical theory
associated to a quiver with ghosts. Nevertheless it indicates that
some of the mathematics used for computing F-term quantities in
physical quivers can be extended to quivers with ghosts.

{\it R-charges and NSVZ beta-function}

For a physical quiver obtained from putting D3-branes at a
singularity, the theory flows to $N = 4$ Yang-Mills theory in the
IR for generic VEVs, but at the origin of moduli space we expect
an interesting $N=1$ CFT. One may try to compute the R-charges of
the fields in the IR by setting the numerator of the the NSVZ beta
functions for the gauge couplings to zero. Typically one does not
find enough constraints, and one employs the strategy of
$a$-maximisation to find the correct $R$-charges as well as the
value of the $a$-anomaly in the IR. Here we will try a similar
procedure for the quiver with ghosts of figure
\ref{quadghostquiver}. For Seiberg dual theories the gauge
invariant chiral operators should identical. Here we also expect
to find the correct value of $a$ as well as the correct
$R$-charges for gauge invariant chiral operators when compared to
a physical quiver for ${\bf P}^1 \times {\bf P}^1$, such as in
figure \ref{P1xP1Ancestor}B. By gauge invariance we mean both the
gauge invariance associated to the nodes as well as the parabolic
symmetries.

The numerator of the NSVZ beta function is
\be
\beta_i = {3} C_2(G) - \sum_{\rm charged\ chirals} (1-2\gamma_i) T(R_i)
\ee
For $SU(N)$ the second Casimir is $C_2(G) = N$ and the index of
the fundamental representation is $T = 1/2$. Moreover at the
conformal point the superconformal algebra relates the $R$-charge
and the dimension of an operator as $\Delta = 1 + \gamma = 3R/2$.
By the symmetry of the quiver and superpotential we expect that
$R_{X_{12}} = R_{X_{34}}$ and $R_{X_{13}} = R_{X_{24}}$. Because
of the symmetries $\delta X_{13} =-  X_{12} X_{23}, \delta X_{24}
= X_{23} X_{34}$ we also get $R_{13} = R_{12} + R_{23}, R_{24} =
R_{23} + R_{34}$. Finally because of this symmetry we know that
the Yukawa couplings must be identical, and we expect they have
dimension zero.

We still need to specify how include the contributions from the
fields of ghost number zero to the beta function. One can think of
$X_{23}$ as giving a contribution to the chiral fields, but
because of its ghost number it has opposite statistics, and thus
the loop integral which calculates its contribution to the beta
function has the same magnitude but the opposite sign from a
normal chiral field.

For node 1 one finds
\be 3N - {1\over 2}(2N(3-2\Delta_{X_{12}}) + 6N^2(3-2\Delta_{X_{41}}) +
4N (3-2\Delta_{X_{13}})) = 0.
\ee
and by symmetry we get the same equation for node 4.
Since the superpotential has R-charge 2 we can solve for $R_{41}$ in
$R_{41} + R_{12} + R_{24} = R_{41} + R_{12} + R_{13} = 2$ and substitute. This gives
\be
 R_{13} + 2 R_{12} = 1.
 \ee
Next we consider nodes 2 and 3 (which will give identical equations):
\be\label{betatwo} 3N - {1\over 2}(2N(3-2\Delta_{12}) + 4 N
(3-2\Delta_{24}) -2N (3-2\Delta_{23}))=0.
\ee
Note we have reversed the sign in the contribution for the ghost.
Using the previous equations $R_{13} = R_{12} + R_{23}$ and
$R_{13} + 2 R_{12} = 1$, we find that (\ref{betatwo}) vanishes
identically and imposes no new constraint. So we have a
1-parameter family of allowed R-charges, parametrised say by
$R_{12}$.

We can compute the 't Hooft anomaly Trace$(R)$. We get
\be
 4N^2 + 4N^2 (R_{12}-1) + 8N^2 (-2R_{12}) + 6N^2 (R_{12}) -
2N^2(-3R_{12})
\ee
which sums up to zero exactly. The first $4N^2$
is the contribution of the gauginos associated with the four
nodes.

Next we will use the proposal by Intriligator and Wecht
\cite{Intriligator:2003jj} and maximise the $a$-anomaly. This
yields $R_{12} = {1\over 2}$. We then have the following table for
the R-charges:
\be
 \matrix{ X_{12} && X_{34} && X_{13} && X_{24} && X_{41} && X_{23} \cr
{1\over 2} && {1\over 2} && 0 && 0 && {3\over 2} && -{1\over 2} \cr }
\ee
Again we are not bothered by the fact that some of the $R$-charges in this table
are zero or negative. The only criterion is that the gauge invariant operators
(the baryons and mesons) have positive R-charge and
dimension, and this includes invariance under the parabolic symmetry.
Moreover, plugging into the
$a$-anomaly, we get
\be
 a={3\over 32}(3{\rm Tr}(R^3) - {\rm Tr}(R)) = {27N^2\over
32}
\ee
which is {exactly} the right answer.

{\it Dibaryon counting}

Another check on the R-charges comes from counting dibaryons
\cite{Witten:1998xy,Gukov:1998kn,Berenstein:2002ke,Beasley:2001zp,Beasley:2002xv,Herzog:2003wt,Herzog:2003dj}.
There should be a 1-1 correspondence between dibaryons of R-charge
$2dN/8$ and curves of degree $d$ on ${\bf P}^1 \times {\bf P}^1$.

Denote by $H_1$ and $H_2$ the homology classes for the left and
right ${\bf P}^1$ respectively. The degree of a curve is given by
intersecting its class with $2H_1 + 2H_2$ using the relations
$H_1^2 = H_2^2 = 0$ and $H_1 \cdot H_2 = 1$. The lowest degree
rational curves are given by an equation $a_\alpha z^\alpha=0$ or
$b_{\dot{\beta}} w^{\dot{\beta}}=0$ and have homology class $H_2$
or $H_1$ and degree 2. The moduli space of such curves is given by
${\bf P}^1$.

Similarly on the quiver side the dibaryons we can write down have $R$-charge at least
$N/2$.
We can construct baryonic operators of $R$-charge $N/2$ as follows.
There is one set which we can make out of $X_{12}$ or $X_{34}$. Recall however
from our discussion of the moduli space that $\del W/\del X_{41}=0$
implies $X_{12} \sim X_{34}$, so we can forget about $X_{34}$ because it will
not give any new operators. Then we can make dibaryons out of $A_1,A_2$ of
the schematic form
\be
(A_1)^s (A_2)^{N-s}
\ee
for $0\leq s\leq N$. This matches with the fact that a D3-brane
wrapped on $H_2 \ltimes S^1$ behaves as an electric particle on
the moduli space of the curve $H_2$, which is ${\bf P}^1$ with $N$
units of magnetic flux. Quantising this moduli space, we find the
$N+1$ sections of $\cO(N)$ \cite{Witten:1998xy}. These dibaryons
do indeed have charge $N/2$.

Similarly one can construct 8 operators that are  invariant  under
the parabolic symmetry and of ghost number one of the form $X_{12}
X_{24}  + X_{13} X_{34}$. From the superpotential we get 6
relations between them, so there are 2 independent such operators.
Then just as with $A_1,A_2$ we can construct $N+1$ dibaryons out
of them with R-charge $N/2$. These presumably correspond to the
states obtained from quantising the moduli space of D3-branes
wrapped on ${\bf P}^1 \ltimes S^1$ where the ${\bf P}^1$ has
homology class $H_1$.

It might be interesting to check some more curves of higher degrees.

\newsubsection{Del Pezzo 3}

%%%%
 \begin{figure}[th]
\begin{center}
        %\resizebox{\textwidth}{!}{
            \scalebox{.5}{
               \includegraphics[width=\textwidth]{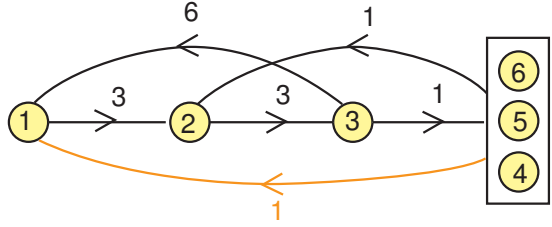}
               }
\end{center}
\vspace{-.5cm} \caption{ \it Quiver diagram for
(\ref{easyDP3}).}\label{DP3ghosts}
 \end{figure}

\begin{figure}[th]
\begin{center}
        %\resizebox{\textwidth}{!}{
            \scalebox{1.0}{
               \includegraphics[width=\textwidth]{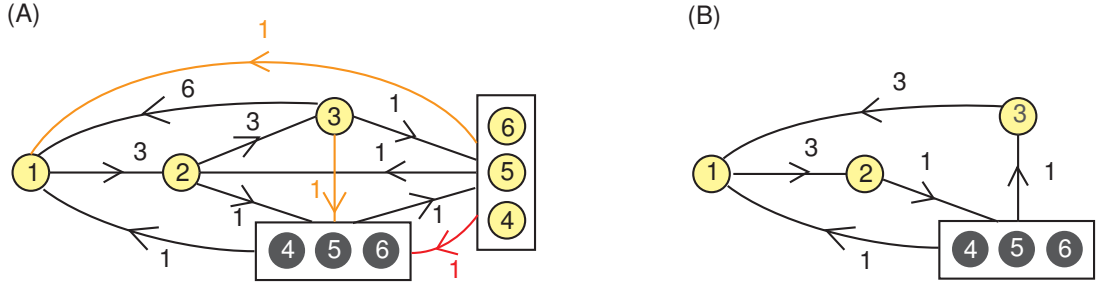}
               }
\end{center}
\vspace{-.5cm} \caption{ \it (A): Quiver for (\ref{extDP3}),
obtained from figure \ref{DP3ghosts} by adding `anti-branes.' (B)
Mutated quiver diagram, which is physical (has no red
lines).}\label{DP3ancestor}
 \end{figure}

Let us consider the case of Del Pezzo 3. As before we will encode the NC structure
through the tensor $f_{ijk}$.  We choose the following
strong exceptional collection
\be\label{easyDP3}
\begin{array}{llll}
1.\ \cO(0) \qquad& 2.\  \cO(H) \qquad & 3.\  \cO(2H) \qquad & 4.\   \cO_{E_1}  \eol
           &              &                 & 5.\ \cO_{E_2} \eol
           &            &               & 6.\ \cO_{E_3}
\end{array}
\ee
The quiver is drawn in Figure \ref{DP3ancestor}A.
We will denote the maps as follows:
\begin{equation}
\begin{array}{ll}
X_{12} = A_i z^i \qquad &
X_{3,456} = D_i\, {\rm res}_{E_i}\eol
X_{23} = B_i z^i   \qquad &
X_{2,456} = E^*_{i}\, {\rm res}_{E_i}  \eol
X_{13} = C^*_{ij}z^i z^j  \qquad &
X_{1,456} = F^*_i \,{\rm res}_{E_i}  \eol
\end{array}
\end{equation}
Here ${\rm res}_{E_i}$ means ``restriction to $E_i$,'' and as usual we will kill three of the nine components
of $(C^*_{ij})^* = C^{ij}$ by adding Lagrange multipliers $Z^k$ and the mass terms $f_{ijk} C^{ij} Z^k$.
Assuming the $E_i$ are exceptional curves obtained from blowing up the points $p,q,r$ respectively,
we find the following superpotential
\be
W = W_{{\bf P}^2}  + (p^\sigma)^i B_i D_1 E^1 + (q^\sigma)^i B_i D_2 E^2 + (r^\sigma)^i B_i D_3 E^3
\ee
with
\ba
W_{{\bf P}^2} &=& \ A_i B_j C^{ij} + f_{ijk} C^{ij} Z^k  .
\ea
Moreover the $F_i$ correspond to ghosts, so we get the following relations:
\ba
\delta E_1 = p^i  F_1 A_i & \qquad \delta C^{ij} = - p^i (p^\sigma)^j  D_1 F_1   \eol
\delta E_2 = q^i  F_2 A_i &\qquad \delta C^{ij} = - q^i (q^\sigma)^j  D_2 F_2 \eol
\delta E_3 = r^i F_3 A_i  &\qquad \delta C^{ij} = - r^i (r^\sigma)^j  D_3 F_3
 \eol
\ea
As one can check, the superpotential is indeed invariant under these symmetries.
Clearly one can write down a very similar quiver theory for any of the Del Pezzo surfaces.

%%%%

Next we would like to do get rid of the ghosts by applying mutations. In the present case it can
be accomplished by shifting $\cO(2H)$ to the right. Then we get a new sheaf
$\tilde{F}_{(3)} = \cO(2H - E_1 - E_2 - E_3)$. The charge vectors are related by
\be
{\rm ch}(\tilde{F}_{(3)}) = {\rm ch}(\cO(2H)) -\sum_{i=1}^3
{\rm ch}(\cO_{E_i}).
\ee
In order to obtain the superpotential for the dual quiver, we first construct the intermediate
quiver by adding the antibranes $\cO_{E_i}[0]$:
\be\label{extDP3}
\begin{array}{ccccccc}
          &   &              &      &  {\cO(2H)[0]}                  &      &                               \eol
\cO(0)[-2]&  \to & \cO(H)[-1]& \to &  \oplus                          & \to & \bigoplus_{i=1}^3 \cO_{E_i}[1]  \eol
           &     &            &     & {\bigoplus_{i=1}^3 \cO_{E_i}[0]}&     &
\end{array}
\ee
The resulting quiver is drawn in  Figure \ref{DP3ancestor}B. We have the additional maps
\ba
X_{\overline{456},1} = \tilde{F}_i &\qquad X_{\bar{4}4} = U_1 \eol  %G_i
X_{2,\overline{456}}= \tilde{E}_i  &\qquad X_{\bar{5}5} = U_2  \eol  %H_i
X_{3,\overline{456}} =\tilde{D}_i  &\qquad  X_{\bar{6}6} = U_3   %K_i
\ea
The composition of maps can be easily  read off from the sheaves. However our main point is that even
if one didn't know the sheaves, it would still be straightforward to read of the compositions of maps of the extended quiver
from the original quiver, and hence find the extended superpotential and symmetries. Clearly there is a correspondence
\be
D_i \leftrightarrow\tilde{D}_i, \quad E_i \leftrightarrow \tilde{E}_i,
\quad F_i \leftrightarrow \tilde{F}_i
\ee
The new superpotential is \ba W &=& W_{{\bf P}^2}  + (p^\sigma)^i
B_i D_1 E^1 + (q^\sigma)^i B_i D_2 E^2 + (r^\sigma)^i B_i D_3 E^3
\eol
 & & +  \tilde{E}_1 U_1 E_1 +  \tilde{E}_2 U_2 E_2 +  \tilde{E}_3 U_3 E_3
+ p^i A_i \tilde{E}_1 \tilde{F}_1 + q^i A_i \tilde{E}_2 \tilde{F}_2
+ r^i A_i \tilde{E}_3 \tilde{F}_3 \eol
\ea
and the new symmetries are
\ba
\delta E_1 = p^i {F}_1 A_i  & \quad \delta C^{ij} = - p^i (p^\sigma)^j  D_1 {F}_1
& \quad \delta \tilde{F}_1 = - U_1{F}_1   \eol
\delta E_2 = q^i {F}_2 A_i& \quad  \delta C^{ij} = - q^i (q^\sigma)^j  D_2 {F}_2
&  \quad\delta \tilde{F}_2 = - U_2 {F}_2
 \eol
\delta E_3 = r^i {F}_3 A_i  & \quad \delta C^{ij} = - r^i (r^\sigma)^j  D_3 {F}_3
& \quad \delta \tilde{F}_3 = - U_3\bar{F}_3  \eol
\ea
and
\ba \delta D_1 =  \tilde{D}_1 U_1  & \quad \delta \tilde{E}_1 =
-(p^\sigma)^i B_i \tilde{D}_1 & \quad  \delta C^{ij} =  p^i
(p^\sigma)^j \tilde{D}_1 \tilde{F}_1   \eol \quad \delta D_2 =
\tilde{D}_2 U_2  & \quad \delta \tilde{E}_2 = -(q^\sigma)^i B_i
\tilde{D}_2 & \quad \delta C^{ij} =  q^i (q^\sigma)^j \tilde{D}_2
\tilde{F}_2\eol \delta D_3 =  \tilde{D}_3 U_3 & \quad \delta
\tilde{E}_3 = -(r^\sigma)^i B_i \tilde{D}_3  & \quad \delta C^{ij}
=  r^i (r^\sigma)^j \tilde{D}_3 \tilde{F}_3 \ea
As in the previous example all this information can be easily lifted from the original
quiver diagram:
\begin{itemize}
\item The superpotential term $(p^\sigma)^i B_i D_1 E^1$,
gives rise to the symmetry $\delta \tilde{E}_1 = -(p^\sigma)^i B_i \tilde{D}_1$.
\item The symmetry $\delta C^{ij} = - p^i (p^\sigma)^j  D_1 {F}_1$
yields the new symmetry $\delta C^{ij} =  p^i (p^\sigma)^j \tilde{D}_1 \tilde{F}_1 $.
\item The symmetry $\delta E_1 = p^i  {F}_1 A_i$
gives rise to the superpotential term $p^i A_i \tilde{E}_1 \tilde{F}_1$.
\item For the new compositions, we add to the superpotential the cubic term
$\tilde{E}_1 U_1 E_1$, and we add the symmetries
$\delta \tilde{F}_1 = - U_1{F}_1,\  \delta D_1 =  \tilde{D}_1 U_1$.
\end{itemize}
Finally there are $X_{4\bar{4}},X_{5\bar{5}},X_{6\bar{6}}$ of
ghost number minus one which parametrise certain redundancies
among the shift symmetries. They are drawn in red.

To get the mutated quiver, we turn on VEVs for $D_1,D_2,D_3$. The
precise expectation value is not important, so we will just set
$\vev{D_i}=1$. Then the ghost number zero fields $\tilde{D}_i$
cancel with the ghost number -1 fields. When turning on the $D_i$,
we get quadratic terms for $B_i$ and $E_i$, so we should solve for
their equations of motion and substitute back. All in all then we
are left with the  quiver diagram in Figure \ref{DP3ancestor}B,
with superpotential
\ba\label{DP3quartic}
W &=& f_{ijk} C^{ij} Z^k+
p^i A_i \tilde{E}_1 \tilde{F}_1 + q^i A_i \tilde{E}_2 \tilde{F}_2
+ r^i A_i \tilde{E}_3 \tilde{F}_3 \eol
& &
- A_1 \tilde{E}_1 U_1 (m^{11} C^{11}+m^{12} C^{12}+m^{13} C^{13} )\eol
& &
- A_1 \tilde{E}_2 U_2 (m^{21} C^{11}+ m^{22} C^{12}+m^{23} C^{13}  ) \eol
& & - A_1 \tilde{E}_3 U_3(m^{31} C^{11} +m^{32} C^{12}+m^{33} C^{13}) \eol
& &
- A_2 \tilde{E}_1 U_1 (m^{11} C^{21} +m^{12} C^{22}+m^{13} C^{23}) \eol
& & -A_2 \tilde{E}_2 U_2(m^{21} C^{21} + m^{22}C^{22}
+m^{23}C^{23}) \eol
& &
- A_2 \tilde{E}_3 U_3(m^{31}C^{21} +m^{32}C^{22} +m^{33}C^{23}) \eol
& & -A_3 \tilde{E}_1 U_1(
m^{11}C^{31}+m^{12}C^{32}
+m^{13}C^{33}) \eol
& &
-A_3 \tilde{E}_2 U_2(
 m^{21}C^{31} +m^{22}C^{32} + m^{23}C^{33}) \eol
 & &
-A_3 \tilde{E}_3 U_3( m^{31}C^{31}+m^{32}C^{32} + m^{33} C^{33})
\eol
\ea
Here
\be
m_{ij} =
\left(%
\begin{array}{ccc}
  p^\sigma_1 & p^\sigma_2 & p^\sigma_3 \\
  q^\sigma_1 & q^\sigma_2 & q^\sigma_3 \\
  r^\sigma_1 & r^\sigma_2 & r^\sigma_3 \\
\end{array}%
\right)_{ij}
\ee
and $m^{ij}$ is its inverse. Recall that $p^\sigma$ is defined to be the
unique vector in the kernel of $f_{ijk} p^i$,
and similarly for $q^\sigma,\ r^\sigma$.

This superpotential is still invariant under the remnant symmetry
\ba
\quad \delta C^{ij} = - p^i (p^\sigma)^j   {F}_1 & \quad \delta \tilde{F}_1 = - U_1 {F}_1   \eol
\quad  \delta C^{ij} = - q^i (q^\sigma)^j   {F}_2  &  \quad\delta \tilde{F}_2 = - U_2 {F}_2
 \eol
 \quad \delta C^{ij} = - r^i (r^\sigma)^j   {F}_3   & \quad \delta \tilde{F}_3 = - U_3 {F}_3
\ea
which kills three components of $C^{ij}$. Once again we can take care of this by adding three Lagrange multipliers
and  the mass terms
\be\label{massterms}
 \bar{p}_i \bar{p}^\sigma_j C^{ij} Y_1 +\bar{q}_i \bar{q}^\sigma_j C^{ij} Y_2 +\bar{r}_i \bar{r}^\sigma_j C^{ij} Y_3
\ee
to the superpotential (\ref{DP3quartic}). The total
superpotential, given by (\ref{DP3quartic}) plus
(\ref{massterms}), is then our final answer for the physical
quiver given in figure \ref{DP3ancestor}B.

We can go one step further and do an additional mutation, to get the exceptional collection (\ref{dpthreeexc})
we studied previously. This mutation actually yields a Seiberg duality on node 2, which was to be expected
because we are now mapping a physical ghost-free quiver into another physical quiver. To carry out the Seiberg duality,
we introduce the meson fields $L_i = A_i \tilde{E}_1, M_i = A_i \tilde{E}_2, N_i = A_i \tilde{E}_3$ and the dual quarks $a^i, h_i$.
We also modify the superpotential to
\ba
W \is f_{ijk} C^{ij} Z^k+
\bar{p}_i \bar{p}^\sigma_j C^{ij} Y_1 +\bar{q}_i \bar{q}^\sigma_j C^{ij} Y_2 +\bar{r}_i \bar{r}^\sigma_j C^{ij} Y_3 \eol
& &+
p^i L_i \tilde{F}_1 + q^i M_i \tilde{F}_2 + r^i N_i \tilde{F}_3
+ h_1 a^i L_i + h_2 a^i M_i + h_3 a^i N_i
\eol
& &
- L_1 U_1 (m^{11} C^{11}+m^{12} C^{12}+m^{13} C^{13} )
- M_1 U_2 (m^{21} C^{11}+ m^{22} C^{12}+m^{23} C^{13}  ) \eol
& & - N_1 U_3(m^{31} C^{11} +m^{32} C^{12}+m^{33} C^{13})
- L_2 U_1 (m^{11} C^{21} +m^{12} C^{22}+m^{13} C^{23}) \eol
& & -M_2 U_2(m^{21} C^{21} + m^{22}C^{22}
+m^{23}C^{23})
- N_2 U_3(m^{31}C^{21} +m^{32}C^{22} +m^{33}C^{23}) \eol
& & -L_3 U_1(
m^{11}C^{31}+m^{12}C^{32}
+m^{13}C^{33})
-M_3 U_2(
 m^{21}C^{31} +m^{22}C^{32} + m^{23}C^{33}) \eol
 & &
-N_3 U_3( m^{31}C^{31}+m^{32}C^{32} + m^{33} C^{33})
\eol
\ea
After some simple field redefinitions this superpotential
coincides exactly with the superpotential we obtained previously
(\ref{dp3pot}).

\bigskip

\noindent {\it Acknowledgements:}

It is a pleasure to thank S.~Benvenuti, F.~Cachazo, S.~Franco,
 N.~Itzhaki, S.~Katz, T.~Nevins, L.~Rastelli, C.~Vafa and
 H.~Verlinde for discussions. This material is  based upon
 work  supported by the National Science Foundation Grant
No. PHY-0243680. Any opinions, findings, and conclusions or
recommendations expressed in this material are those of the
authors and do not necessarily reflect the views of the National
Science Foundation.

\end{document}